\documentclass[12pt]{article}
\usepackage[cp866]{inputenc}
\usepackage[russian]{babel}
\usepackage{graphicx}
\begin{document}
\addtolength{\topmargin}{-\topmargin}
\addtolength{\topmargin}{.1in}\addtolength{\textheight}{.2\textheight}
\addtolength{\headheight}{-\headheight}\addtolength{\headsep}{-\headsep}
\oddsidemargin 0.cm\hsize 160mm
\raggedbottom\renewcommand{\baselinestretch}{1.2}
\newcommand{\rot}{\mathop{\rm rot}\nolimits}
\newcommand{\grad}{\mathop{\rm grad}\nolimits}
\newcommand{\diver}{\mathop{\rm div}\nolimits}
\newcommand{\km}{\mathop{\rm km}\nolimits}
\newcommand{\s}{\mathop{\rm s}\nolimits}
\newcommand{\keV}{\mathop{\rm keV}\nolimits}
\newcommand{\eV}{\mathop{\rm eV}\nolimits}
\newcommand{\Ma}{\mathop{\rm Ma}\nolimits}
\newcommand{\Fr}{\mathop{\rm Fr}\nolimits}
\newcommand{\const}{\mathop{\rm const}\nolimits}
\newcommand{\Erf}{\mathop{\rm Erf}\nolimits}
\newcommand{\Ree}{\mathop{\rm Re}\nolimits}
\normalsize

\centerline{\bf EXACT EXPANSION LAW}

\centerline{\bf FOR RICHTMYER-MESHKOV TURBULENT MIXING ZONE}

N.A. Inogamov$^1$ \& A.M. Oparin$^2$

1. Landau Institute for Theoretical Physics, Moscow

2. Institute for Computer-Aided Design, Moscow

\centerline{Abstract}

{\it Definition of mixing fronts and turbulent mixing zone.}
Development of the Richtmyer-Meshkov instability
leads to mixing of two substances
separated by contact boundary.
Let the axis $z$ is perpendicular to a plane of boundary
and $z=0$ corresponds to an unperturbed plane.
The turbulent mixing zone is limited by front $z=h_+(t)>0$
from heavy fluid side
and by front $z=h_-(t)<0$ from light fluid.
Mixture of two fluids is confined between these two fronts
(layer $h_-<z<h_+).$

{\it Initial conditions.}
Before pass of a shock wave both liquids were still
[were at rest: $\vec v(x,y,z,t<0)\equiv 0$]
and $z=\eta_{ini}(x,y)$ was perturbed contact boundary.
We consider linear (small amplitude) perturbation: $|\nabla\eta_{ini}|\ll 1.$
In this case the shock wave passes perturbations quickly:
$t_s\ll \lambda_{ini}/|\dot\eta_{\lambda ini}|,$
where $t_s$ is shock wave passage (sonic) time,
$\lambda_{ini}$ is a typical wavelength (or space scale) of perturbations,
and $|\dot\eta_{\lambda ini}|$ is typical velocity after shock wave passage.

{\it Transition and asymptotic stages.}
Universal asymptotics
follows the transition stage at late time
$t\gg \lambda/|\dot\eta_{\lambda ini}|.$
Functions $h_\pm(t)$ asymptoticaly transform to power law dependences.
Mixing is accompanied by cascade of enlarging of dominant scale
of a turbulent mixing zone.
Under condition $\mu=0$ exact scaling laws are:
$h_+^{2D}\propto t^{2/5}$ and $h_+^{3D}\propto t^{1/3},$
where $\mu$ is density ratio,
2D and 3D mark dimension of space.
These scalings are calculated in our work.
The scaling exponents 2/5 and 1/3
are determined by the mechanism of redistribution
of $z-$component of momentum from small into the large scales.

For $\mu\neq 0$ the exponents 2/5 and 1/3
define the bottom limit of a range of possible values
for both functions $h_+(t)$ and $h_-(t).$
Our 2D direct numerical simulations at different values $\mu\sim 1$
show that $d\ln h_\pm/dt\approx 2/5.$

{\bf This work has been supported by RBRF (grants 02-02-17499, 03-01-00700)
and scientific schools (NSh-2045.2003.2, NSh-70.2003.1).}

\newpage
\centerline{{\bf \# 1. Introduction}}

\vspace{.5cm}
The Richtmyer-Meshkov instability
starts to develop after crossing (passage) by a shock wave of a contact boundary.
In case of small on amplitude initial perturbation
evolution in time proceeds in three stages
(we shall designate them I, II and III).
At short acoustic stage I\footnote{
In case of small initial perturbations
"hierarchy" of characteristic times for the stages I, II and III
is following:
$t_I\ll t_{II}\ll t_{III}$.
}
formation of a near boundary velocity field completes.
This field appears due to interaction of contact boundary with a shock wave.
Near boundary field defines the further hydrodynamical\footnote{
In case of small initial perturbations
hydrodynamical motion of substance at stages II and III
occur to velocities small in comparison with sonic velocity
(incompressible motion).
At stage I at presence of a shock wave
with Mach number Ma significantly larger than 1
compressibility effects are essential.
} motion.

Duration of an acoustic stage is about
$$
t_s \sim \lambda_{ini}/c_s^{min}. \eqno (1)
$$
Where $\lambda_{ini}$ is typical transverse scale of initial perturbations,
$c_s^{min}$ is minimal from sonic velocities after shocks
or after expansion wave.
In case of usual equation of state this is sonic velocity in heavy gas
because pressure is continuous at contact boundary
and $c_s\sim \sqrt{p/\rho}$.

The expression (1) includes space scale $\lambda_{ini}$
(not amplitude $\eta_{\lambda ini}$)
as acoustic interaction of the curved contact boundary
with the curved shock waves continues
while shock waves will not depart from boundary on the large distances
$>\lambda_{ini}$.
Interaction is carried out by means of a series of oblique sound waves
which move in the expanding layer between shocks and boundary.
This interaction
creates vorticity which defines motion at stages II and III\footnote{
Baroclinic generation of vorticity at the boundary
and generation of vorticity in the bulk by curved shocks
(Schilling et al., 2004; Wouchuk, 2001a, 2001b).
In the case of weak shocks we can neglect bulk vorticity.
}.

At a stage II near boundary velocity field decays
at removal from boundary on distances which exceed $\lambda_{ini}$.
The law of reduction in velocity in depth from boundary
depends on transversal correlations of perturbations.
This law qualitatively differs
for periodic, solitary and random perturbations\footnote{
Exponential decay for periodic perturbations
and power laws for random and solitary perturbations
at the same effective depth $\sim\lambda_{ini}$ of near boundary velocity field.}.
In turn decay of velocity in depth
determines the law of expansion of a zone of turbulent mixing at a stage III.
The given work
is devoted to the analysis of distribution of velocity near boundary.

At stages II and III
entropy and vortical modes on the one hand
and acoustic modes with another - are decoupled\footnote{
They are very different in velocities.
Entropy and vortical modes are slow
while acoustic modes are fast.
}.
Therefore an influence of compressibility becomes nonsignificant.
Duration of sonic stage (1) is small in comparison with main time scale
$$
t_{\lambda ini}\sim \lambda_{ini} /w_{\lambda ini}. \eqno (2)
$$
The time (2) is necessary
for development of perturbations up to nonlinear or "saturated" amplitudes.
Displacement of boundary during short time (1) is small
in comparison with $\lambda_{ini}$.

An estimate based on the Richtmyer formula gives
$$
w_{\lambda ini}\sim
 {\rm At}^+ k_{ini} \eta_{\lambda ini}^+ W_{unpert CB}, \eqno (3)
$$
where At$^+$ is Atwood number after (sign plus) shock,
$k_{ini}=2\pi/\lambda_{ini},$
$\eta_{\lambda ini}^+$ is typical amplitudes of boundary perturbations
(after shock)
when we shift at distance $\sim \lambda_{ini}$ in transverse direction,
$W_{unpert CB}$ is unperturbed velocity of contact after shock,
$W_{unpert CB}\sim c_s.$
From equation (3) follows,
that a smallness of $t_s$ (1)
in comparison with $t_{\lambda ini}$ (2)
it is caused by linearity of initial perturbations.

Let
$$
\eta_{\lambda ini}^- \sim \lambda_{ini} \eqno (4)
$$
(nonlinear initial perturbations),
here $\eta_{\lambda ini}^-$ is amplitude of surface perturbation before a shock.
In this case we cannot separate stages I and II:
$t_s \sim t_{\lambda ini},$
$w_{\lambda ini}\sim W_{unpert CB}\sim c_s.$
Hence, the stage III "will start"
with high initial velocities
making an appreciable share from sonic velocity.
Nevertheless,
received below asymptotical expansion laws
of a turbulent mixing zone
do not vary
in case of large perturbation (4)
in comparison with a case of linear perturbations.
The matter is that
eventually velocities are reduced with time,
and become small in comparison with $c_s$.

Motion at a stage II is caused by movement of the primary vorticies
$\sim\lambda_{ini}$
created by a shock from surface perturbations at a sonic stage.
As we will see,
the essence of motion comes to light
at introduction of concept about generations
and terms primary, secondary, tertiary, etc.
The question is the current generation of a chain
of sign-variable vorticies
of the size $\lambda_{curr}(t)$
and about evolution of this generation.
Such chain refers to also as dominating or large-scale structure.
Evolution consists in growth of the size $\lambda_{curr}(t)$
as a result of merging of vorticies.
Each stage of merging
connects parental generation
and generation of first "descendants" of these "parents".
Asymptotical law of expansion
is established for far "descendants" of primary vorticies.
Motion at a stage II has transitive character and is not universal.

Fundamental interest represents
a finding of the law of expansion
of a zone of mixing at large times $t\gg t_{\lambda ini}$.
This is an asymptotical turbulent stage
(stage III)
of development of the Richtmyer-Meshkov instability.
Appears, present universal asymptotics, describing a stage III.
Its representation makes the basic contents of the given work.
As against laws of mixing
at Rayleigh-Taylor and Kelvin-Helmholts turbulence
when exists simple self-similarity,
here simple self-similarity is not present.
In Rayleigh-Taylor and Kelvin-Helmholts cases
there is a unique dimensional parameter:
free fall acceleration $g$
or a velocity difference $\Delta u$ accordingly.
In the Richtmyer-Meshkov case
there are two independent parameters
$\lambda_{ini}$ and $w_{\lambda ini}$.
Therefore simple self-similarity is absent.
At such position presence universal asymptotics
is rather not trivial bright phenomenon.
The mechanism is below described,
responsible for expansion of Richtmyer-Meshkov turbulent zone
with self-similar hierarchy of generations.

At stage III
during each moment of time
in a mixing layer
the vortical structure dominates
over the current scale $\lambda_{curr}(t) \sim h(t)$,
which grows with $t$
$(h$ is thickness of a layer of interpenetration).
Evolution in time
consists in change of one dominating structures
by others by their enlarging
(the inverse cascade, changing of generations).
Externally
it reminds a situation with Rayleigh-Taylor and Kelvin-Helmholts mixing.
Actually mechanisms of enlarging
is essentially different
in Rayleigh-Taylor and Kelvin-Helmholts case
on the one hand
and Richtmyer-Meshkov with another.

In the first case
there is the plentiful external source
which amplifyes subharmonics of the current dominating scale.
In a Rayleigh-Taylor case it is gravitational energy,
reserved in a heavy liquid.
The heavy liquid is supported above light liquid
due to pressure of light liquid.
Therefore gravitational energy of a heavy liquid
cannot be liberated in free falling.
Downturn of the centre of gravity of system
from heavy and light liquids
occurs by exchanges or rearrangements of heavy and light liquids.
Accordingly Rayleigh-Taylor instability
is called also exchange or interchange instability.
In Kelvin-Helmholts case
a source is kinetic energy of motion
which is external to a turbulent mixing zone.
Alternation of generations
consists in replacement of harmonics of the current dominating scale
(parents)
on subharmonics of these harmonics
(descendants).

In the second (Richtmyer-Meshkov) case
an external source is not present.
As it will be shown in the given work,
fading Richtmyer-Meshkov motion
is supported due to statistical redistribution
of a longitudinal momentum from small scales in large scales.
Thus in the second case (as against the first)
energy of movement in the large scales is less
than energy of movement,
Which has been concentrated in small scales
on the previous stages of development of turbulent mixing.

In \# 2 the statement of a problem is formulated.
Connection between disintegration of break
and Richtmyer-Meshkov instability is considered
(generation of vorticies on contact at disintegration of break).
It is spoken about "processing" of surface perturbation $\eta$
into near boundary velocity non-uniform field $\vec v$.
The example of typical stochastic function
with the finite scale of heterogeneity $\lambda_{ini}$ is presented.
Functions from this class characterize general initial perturbations.

\noindent In \# 3 characteristics of a mixing zone are described:
(i) Average thickness of a mixing zone
in which an interpenetration of phases takes place,
(ii) large-scale structure (sequence of vortical dipoles),
and (iii) small-scale details inside large-scale dipoles.

\noindent In \# 4 value of the pressure
created in a liquid by the slowed down bubble is emphasized.
The important examples
with a periodic chain of bubbles and a solitary bubble
are considered.

\noindent In \# 5 the pressure and velocity fields
near a random chain of bubbles is analysed.
Comparison of periodic, solitary and random arrangements of bubbles
is instructive.

\noindent In \# 6 spectral characteristics of a velocity field
are calculated.
It is found,
that the major long wavelength part of a spectrum
depends on wave number as a power law.

\noindent In \# 7 the velocity field
arising directly after passage of a shock wave
is analysed.

\noindent In \# 8 the law of expansion of a turbulent mixing zone
into a heavy liquid is obtained
$$
(h_+\propto t^{2/5}).
$$

\noindent In \# 9 dependence on time
of large-scale velocity fluctuations is discussed
(changeability with approximately constant step on $\ln t$
in process of change of dominating structures in the inverse cascade).

\noindent In \# 10 the remark on mixing is made
at arbitrary value of Atwood number.

\noindent In \# 11 stimulated and spontaneous turbulence are compared.
In a stimulated case expansion of a turbulent mixing zone
is connected to action of beforehand created large-scale perturbations.
In a spontaneous case mixing develops from perturbations
which have finite initial scale $\sim \lambda_{ini}$.
As against stimulated case
it is a problem about decay of Richtmyer-Meshkov turbulence.
It is interesting to consider motion on large times
When thickness of a mixing layer
is large in comparison with initial scale,
$h_+\gg\lambda_{ini}$.

\noindent In \# 12 received results
are generalized on a three-dimensional case.
The law
$$
h_+\propto t^{1/3}
$$
of expansion of mixing front
is found.

\vspace{.5cm}
\centerline{{\bf \# 2. Statement of problem}}

\vspace{.5cm}
In a problem
about Richtmyer-Meshkov instability
examine the movement
arising owing to passage of a shock
through contact boundary between two different gases.
In fig. 1 straight lines 1 and 2
correspond to contact boundary (a straight line 1)
and a shock wave (a straight line 2).

At the moment $t=0$ the wave reaches boundary.
At $t> 0$ from boundary
in the different sides
depart or two shock waves (a case 1),
or a shock wave and a rarefaction (expansion) wave (a case 2)
(see, for example, reviews in monographies
Inogamov, Demianov, and Son 1999,
Inogamov 1999
and in articles Wouchuk 2001a, 2001b).
What from these two cases is carried out
depends on the density ratio,
thermodynamic properties of gases
and Mach number of a shock.
Usually in case of 1
shock goes from light gas
into heavy gas ("from light to heavy"),
and in case 2 - on the contrary.
Thus the rarefaction wave
leaves from contact boundary, moving on heavy gas.

Problem about a shock wave,
moving through plane contact boundary between two homogeneous semispaces,
is a special case of the well-known important problem about disintegration of break.
At the analysis of Richtmyer-Meshkov instability usually a case with substances,
still before passing of a shock, is considered.
Certainly, generalizations of this analysis are possible
which include others cases of disintegration of break
(generation of vorticity on and near the perturbed contact boundary
at disintegration of general break).
For example, it is interesting to consider a case
in which before occurrence of shock two homogeneous semispaces move towards each other
(Demchenko, 2004).

Richtmyer-Meshkov instability arises
when boundary between gases has perturbations.
In Fig. 2 the example of the perturbed boundary is shown.
The axis $z$ (a longitudinal axis) is directed on a normal to unperturbed boundary.
Below plane (2D) and three-dimensional (3D) spaces are considered.
The coordinate $x$ $(x, y)$ is transverse coordinate in 2D (3D) space.

Let's consider coordinate frame connected to unperturbed boundary.
After passage of a shock through boundary
there is a non-uniform near boundary velocity field $\vec v(\vec r, t)$.
It causes movings of boundary $\eta (x, y, t)$
concerning position of unperturbed boundary
(change of an amplitude of perturbations $\eta$ in time).
In Fig. 3 movement of boundary
in cases in a phase and an counterphase is shown
(see, for example, Inogamov, Demianov, and Son 1999, Inogamov 1999).
We compare here an angular phase of perturbations of boundary $\eta$
with a phase of velocity $\dot\eta$.
Movement in a phase
takes place at passage of a shock from light to heavy gas,
and in an counterphase at passage of a shock from heavy to light gas
(see Fig. 3
on which positive directions of longitudinal axes of coordinate $z$
and velocity $w$ are shown).

In Fig. 4 velocity distribution
$\vec v(x,z=0,t_{fix})$
Or $ \vec v(x, y=y_{fix}, z=0, t_{fix})$
along transverse coordinate $x$ (or $x, y)$ is shown
in 2D (or 3D) case.
As it was told, it is formed as a result of passage of a shock
through the perturbed boundary.
Plane $z=0$ corresponds
to the current position of unperturbed boundary
(dashed straight lines in Fig. 3).
Velocity components $u, v$ and $w$
concern to axes $x, y$ and $z$
(compare Fig. 3 and 2).

Because of the wavy form of the perturbed boundary (Fig. 2)
with scale $\sim\lambda_{ini}$
the near boundary velocity field appears sign-variable
with characteristic scales of wavelength $\lambda_{ini}$
and amplitudes $w_{\lambda ini}. $
Here the index "ini" specifies the initial moment of time.
Let's consider a case of small perturbations
$ \eta_{\lambda ini}^+/\lambda_{ini} \ll 1, $
when Richtmyer formula is applicable.
The scale $ \eta_{\lambda ini}^+ $
is a random mean square deviation of an interface
at shift along boundary on distance $ \sim\lambda_{ini}. $
The index "plus" means
that the given value is taken after passage of a shock.
As it was spoken (\# 1)
found below power law expansion of a turbulent mixing zone
remains fair for nonlinear initial perturbations (4).

With dimensionless factor Richtmyer formula (3) looks like
$$
w_n = F {\rm At}^+ k_n \eta_n^+ W_{unpert CB},\;\;\;
w_{nm} = F {\rm At}^+ k_{nm} \eta_{nm}^+ W_{unpert CB}. \eqno (5)
$$
It is used for recalculation of boundary perturbations $ \eta $
into velocity field
 $ \dot\eta $ or $w. $
In (5) $F $ is dimensionless factor
dependent in case of ideal gases from their adiabatic exponents,
from temperatures of gases,
and Mach number of a shock,
see, for example, monographies
Inogamov, Demianov, and Son 1999, Inogamov 1999
with reviews of articles with calculation of factor $F. $
In an incompressible case we have $F=1. $
This case corresponds to impulsive acceleration $g\propto \delta (t). $
In (5) At$^+ =\frac {1-\mu^+} {1 +\mu^+}, $
$ \mu^+ $ is the density ratio after shock,
$W_{unpert CB} $ is velocity of unperturbed boundary
(a dashed straight line, Figs. 1-3)
after shock
in laboratory frame
in which substances before a shock are in rest.
In case of small perturbations
at a linear stage of development of instability
a principle of linear superposition is valid.
In (5) $k_n=2\pi n/L, $
$n $ is number of Fourier harmonic (2D),
$k_{nm} = \sqrt{(k_n^x)^2 + (k_m^y)^2} $ (3D),
$k_{nm} $ is a wave number of a harmonic with $x $ and $y $ components
$k_n^x, k_m^y $ in 3D geometry.
We put our system into large $(L\gg \lambda_{ini})$ "box"
with transversal sizes $L $ (2D) Or $L\times L $ (3D).
Perturbations continue periodically
outside the end points of the "box".

{\bf  Comparison of amplitude of harmonic
and a deviation due to shift.}
Values $\eta_n^+$ and $\eta_{nm}^+$ are Fourier amplitudes of harmonics.
It is necessary to distinguish amplitudes $ \eta_n, $ $w_n $
of harmonics $k_n $ $ (k_n=2\pi/\lambda_n) $
from random-mean-square deviations due to transverse shift $ \lambda_n $
$$
\eta_{\lambda n}(t)=\sqrt{ \langle [\eta(x,t)-\eta(x+\lambda_n,t)]^2 \rangle_x },
$$$$
w_{\lambda n}(z,t)=\sqrt{ \langle [w(x,z,t)-w(x+\lambda_n,z,t)]^2 \rangle_x },\;\;\;
\langle f \rangle_x = \int_0^L f dx/L. \eqno (6)
$$
Equations (6) can be generalized for 3D case.
The random-mean-square deviation of velocity
depends on velocity distribution
which is formed by addition of many harmonics.
Therefore for random functions $f$
formed by an interference of harmonics $f_n $
with a random phase $ \psi_n $
we receive
$$
f_{\lambda n} \sim \sqrt{n} f_n. \eqno (7)
$$

\vspace{.5cm}
\centerline{{\bf \# 3. Structure of turbulent mixing zone}}

\vspace{.5cm}
{\bf Thickness of mixing layer.}
At $t <0 $ there are two semispaces "h" and "l" in contact (Fig. 5a)
filled with pure substances with density $ \rho_h $ and $ \rho_l. $
Thus a vertical (function of $z$) profile of average concentration
$ \bar c (z, t=0) = \langle c \rangle_{x y} $ (see (6))
of light phase "l"
is the "step"
which is drawn
by dashed straight lines in Fig. 5b.
At moment of time $t=0$ the shock arrives at unperturbed contact boundary.
After that in a picture of distribution of phases
a new important zone appears -
this is the turbulent mixing layer M (Fig. 5c),
filled with mixture of substances "h" and "l".
The quantity $c (\vec r, t) $
gives a volume fraction of light phase in an element of volume.
Density of mixture
which fills this volume
is equal
$$
 \rho_{mix} (\vec r, t) =c\rho_l + (1-c) \rho_h.
$$
After appearance of mixing zone
the profile $ \bar c (z, t) $ "is smeared"
(Fig. 5b, a continuous curve)
on thickness of a layer of mixing $h=h_+ + h_-. $
Coordinates $z=h_+ $ and $z=h_-$
(last points of interpenetration of phases)
correspond to fronts of mixing:
$h_+ $ is a front of bubbles, and $h_-$ is front of jets
$ (\rho_l <\rho_h). $

{\bf Transverse large-scale structures.}
Let us consider space structure of mixing zone.
The one-dimensional profile of average concentration $\bar c(z,t_{fix})$
(Fig. 5b)
reminds diffusive profiles.
Intermolecular distances and lengths of free paths of molecules
are small in comparison with thickness $h.$
Therefore molecular diffusive profiles are really one-dimensional.
In a turbulent case
magnitudes of the turbulent fluxes
carrying phases along the $z$ axis
and rate of expansion of a mixing zone
are defined by large-scale heterogeneities of mixing zone.
It is important
that the transverse sizes $\lambda_{curr}$ of these heterogeneities are large
$$
\lambda_{curr}\sim h_+. \eqno (8)
$$

Average flux $\langle c w \rangle_{x y}$ of light phase "upwards" (to $z>0$)
and a heavy phase "downwards" $\langle (1-c) w \rangle_{x y}$
are opposite.
The mixing zone M (Fig. 5c) is two-stream region of counter motion of phases.
Transport of "l" and "h" substances
occurs by interchangement of volumes "l" and "h":
h$\leftrightarrow$l (Fig. 6).
Usually interchanges are connected with Rayleigh-Taylor instability
where they free gravitational energy.
Here (Richtmyer-Meshkov instability)
interchanges (Fig. 6) occur because of inertia of motion,
as the only possible form of motion.
Certainly, any movings of liquid volume
to an incompressible liquid are exchange motion.

In space large-scale (Fig. 6) two counter stream motion
is carried out through alternation of jets, "fingers" or columns
of light "l" and heavy "h" liquids (Fig. 7a).
This alternation
causes changes of a sign on vertical velocity $w$
as function of transverse or horizontal coordinates (Fig. 7b).
Conformity of columns and signs on velocity
is underlined by the arrows
connecting Figs. 7a and 7b.
The correlator $K_{c w}(z,t)=\langle (c-1/2) w \rangle_{x y})$
is positive.
The current thickness $h$ of a mixing layer
is defined on positions of last points
$h_{\pm} $ (Fig. 7a).
Horizontal arrows show designing these points on an axis $z. $

Sign alternation of velocity and alternation of columns
are caused by a chain of vortical "spots" (Fig. 8a)
in which vorticity turns off.
Signs on circulation of "spots" in a chain alternate.
It is shown in Fig. 8a with the help of the bent arrows in circles
and in Fig. 8b by signs plus and a minus in circles.
Vortical pairs (dipoles)
involve in a zone of mixing pure "l" and "h" liquids
(arrows "l" and "h" in Fig. 8a).

Base element of large-scale structure
is the vortical pair or a vortical dipole (Fig. 8) of the size (8).
The pair the next columns is connected to each dipole.
Occurrence of vorticities
occurs because of a variation of a sign on the momentum
transferred by a shock to a near boundary layer of substance.
In turn the variation of a momentum
is caused by wagging about average
the perturbed contact boundary $ \eta^-(x, y). $
Vortical circulation is connected
to the angular momentum of liquid.

Vortical motion at a stage III (\# 1)
is described in approximation of an incompressible liquid
(conservation of volume).
Thus overflows upwards should be compensated
(Fig. 6)
by returning of equal volume of a liquid downwards:
$ \int_0^L \int_0^L w (x, y, z=0, t) dx dy =0. $
At random perturbations
transverse correlations between base elements (between large-scale dipoles)
fade on scale $\lambda_{curr}$ (8).
That base elements
appear under the order of size of one size
along all infinite chain
is caused by transversal uniformity
on the average
of initial perturbations $ \eta^-. $
Therefore compensation of fluxes of volume
$ \int\int w dx dy $
occurs mainly at size (8).
Necessity of compensation
results to interchanges (Fig. 6) and to alternation of columns (fig. 7).

{\bf Previous stages and fractal structure.}
The instant spatial picture is above submitted.
Its basis is the large-scale chain of the dipoles set in scales
$\lambda$ (8) ш $w_{rms},$
where
$$
w_{rms}(t)= [ \int_0^L \int_0^L [w(x,y,z=0,t)]^2 dx dy/L^2  ]^{1/2}. \eqno (9)
$$
Amplitudes of fluctuations (9)
is of order of random mean square
$w_{\lambda}$ (6) at dipole size.

Let this instant picture concerns to the moment of time $t^i. $
The index $i =\ln (t/t_{II}) $ (2)
runs consecutive stages of development of the inverse cascade
(sequence of enlarging of scale $ \lambda^i). $
At each enlarging the index $i $ changes on size $ \Delta i\sim 1. $
In Fig. 9 two large-scale chains of dipoles $ \lambda^i $ at stages $i_1 $
and $i_2 $ $ (i_2-i_1\gg 1)$
are shown.
At a stage $t^{i2} $ in a dipole $a_2 $
there are almost all dipoles $a_1 $ from a stage $t^{i1} $
got in an interval $x_{a2} <x <x_{a2} + \lambda^{i2}. $
Thus, the dipole $a_2 $ is the claster from the specified dipoles $a_1. $

In Fig. 10a dipoles of two consecutive generations
$i_1 $ and $i_2 = i_1 + \Delta i $ are shown.
In Fig. 10b three generations
$i_1, $ $i_2 = i_1 + \Delta i $ and $i_3 = i_2 + \Delta i $
are shown.
Large-scale vortical rotation strongly deforms and stretches
small-scale details connected to the previous generations (Fig. 11).
The analysis of numerical calculations shows
that these details of current dominating structure
are more likely relicts of previous generations
than result of Kolmogorov (3D) or Kraichnan fragmentations.
In such hierarchical way
internal fractal structure of large-scale formations forms.
This fractal
is responsible for rate of mixing of phases at a molecular level,
see Redondo and Garzon, 2004.

After separation molecular mixture remains as an intermediate layer
in experiments on separation with change of a sign on acceleration,
Kucherenko et al., 2004a.
The contents of light phase
in unit of volume of a mixture miscellaneous (varies from 0 up to 1).
Studying an intermediate layer
it is possible to establish
what share of the given layer
having this or that contents of light phase.

To estimate a share of molecular mixed mixture
in total amount of mixture
in turbulent zone
it is possible by use of chemical indicators
(Linden, Redondo, Youngs, 1994;
Kucherenko et al., 2004b).
Recent work (Chertkov, 2003)
is devoted to studies of process of turbulent fragmentation
during development of Rayleigh-Taylor mixing.

\vspace{.5cm}
\centerline{{\bf \# 4. Pressures and momentum, At$=1$}}

\vspace{.5cm}
Let's describe calculation of an exponent $ \theta $
in a case $ \mu=0, $ At $ = 1 $ (mixing with "vacuum").
At a stage III flow tends to universal (self-similar) regime (\# 1)
in which $h_+\propto t^{\theta}. $
Calculation $ \theta $ is based on the law of momentum conservation.
Transfer of momentum to liquid
is carried out due to ram pressure $ \rho v^2 $
of bubbles "imprinting" into liquid.
Therefore creation of momentum by pressure is interesting for us.

{\bf Pressure distribution. Periodical case.}
Let's consider a periodic chain of bubbles.
Periodicity means that all bubbles in a chain are identical.
Bubbles are symmetric concerning the vertical average line
(b$\infty), $
see Fig. 12b.
The important point b is an apex of a bubble.
In this figure the symmetric half of period of flow is shown.
To receive all flow
it is necessary to complete symmetrically to the left
half of bubble
up to the whole bubble.
The full chain is obtained by periodic continuation
of the whole bubble to the left and to the right\footnote{
About periodic solutions in 2D and 3D spaces see, for example,
reviews Inogamov, Demianov, and Son 1999;
Inogamov 1999;
and works Inogamov and Oparin, 1999;
Inogamov et al., 2001a.
Here for simplicity the example in 2D a case is presented.}.

Velocities of lagrangian particles
on a straight line (b$\infty) $
are directed upwards (to $z> 0), $
arrows B in Fig. 12a and b,
"imprint" of bubbles.
These velocities (in the straight b$\infty) $
decrease in due course
(deceleration of bubbles, inertial imprint of bubbles).
Hence,
on the straight line (b$\infty) $
the vector $ \nabla p $
is directed upwards.
Distribution of pressure
$p (x=x_b, z, t_{fix}) $ on an axis $z $
is shown in Fig. 12a,
$x_b $ is coordinate of the bubble apex b.
In case of an incompressible liquid
the reference mark of pressure can be any function of time.
In region occupied by "vacuum"
(the region below the boundary of bubble bj, Fig. 12b)
pressure is homogeneous:
$p (\vec r, t) \equiv p_{CB} (t). $
Let's put
$p_{CB} \equiv 0 $
(a free choice of "calibration" of pressure).
Then function $p(z) $ in Fig. 12a begins from zero $p(z=z_b) =0. $

Function $p(z) $ in Fig. 12a
grows with growth $z$
(as it is told, $dp/dz> 0). $
Asymptotically
at the large removals from roughnesses of boundary bj
$ (z\gg\lambda) $
heterogeneity of pressure fades -
pressure becomes homogeneous:
$$
p\to p_{\infty}(t)> p _ {CB} =0,
$$
above $ \lambda $ is the period of a chain of bubbles (Fig. 12b).
It is shown in Fig. 12a and c.

In a liquid
the gradient $ \nabla p $
is finite on boundary bj between liquid and "vacuum".
In jets j (liquid deformations aside "vacuum")
the gradient $ |\nabla p| $ is small.
Vertical straight line (j$\infty) $ is a middle line of a jet.
In Fig. 12b the family of instant isobars 0, 1, 2 and 3 is presented.
Pressure accrues with growth of number of an isobar.
Zero isobar is the boundary bj.

Initial kinetic energy of motion of liquid
gradually passes in kinetic energy of jets.
The momentum
$$
 \int_0^t p_{\infty} dt 
$$
passes into a momentum of the substance
which have been "thrown out" in jets j.
Let's result an example.
Let the condition with plane boundary
and one harmonic of velocity was initial:
$$
z =\eta (x, t=0) \equiv 0, \;\;\;
\vec v = \nabla \varphi, \;\;
\varphi (x, z, t=0) = - (w_{ini}/k) e^{-kz} \cos kx,\;\;z> 0.
$$
Then initial vertical $ (w =\varphi_z) $ momentum
 $ \int_{-\infty}^{\infty} dx \int_0^{\infty} w dz $
is equal to zero.
At $t=0 $ velocities of apex points b
$ (w_b, x_b=0) $
 and j $ (w_j, x_j =\pi/k) $
are equal each other $ \pm w_{ini}. $
At $t\to\infty $ absolute value of velocity $w_j $
is approximately twice larger than $w_{ini},$
Inogamov et al., 2001a.
Integral
$$
I(t) =
\lim_{L\to\infty}
| \int_0^L (dx/L) \int_{\eta (x, t)}^{\infty} w dz |
$$
(where function $ \eta (x, t) $
gives the current position of contact boundary)
accrues in due course $t $
and tends to limiting value
$$
 \int_0^{\infty} p_{\infty}/\rho_h dt. 
$$

So, first, in a periodic case we have
$p_{\infty}> p_{CB}$
(the case $ \mu=0$ is considered),
$$
p_{\infty}-p_{CB}\sim \rho_h w_b^2. \eqno (10)
$$
Second, heterogeneity of pressure are connected to roughnesses of boundary.

Let's note that at $ \mu\neq 0 $ "hole" or depression in pressure is connected
with compact region of vortical mixing
(pressure in vortical region is lower than $p_{\infty}). $
Let's consider mixing near a horizontal plane $z=0 $ (3D)
or a horizontal straight line $z=0 $ (2D).
When plane mixing zone is present (Fig. 13a)
then in distribution of pressure "valley" (2D)
extended along an axis $x$ is formed.
Average $ \langle\rangle_x $ (6)
instant profile of pressure $p(z) $ is shown in Fig. 13b.
Differences of pressure
$\Delta p_{\infty}^h = p_{\infty}^h - p_{min}\sim \rho_h w_b^2$
and $\Delta p_{\infty}^l\sim \rho_l w_j^2$
correspond to two semispaces $z\to\pm\infty $
filled still pure "h" and "l" liquids
with densities $ \rho_h, $ $ \rho_l, $
$p_{min} $ is pressure in a minimum.
At a stage $t\sim t_{II} $ (2) we have $w_b\sim w_j\sim w_{ini}. $
Accordingly $ \Delta p_{\infty}^h> \Delta p_{\infty}^l $
as $ \mu <1, $ $ \mu =\rho_l/\rho_h. $

{\bf Pressure distribution. Solitary case.}
Let's consider a case
when into liquid semispace $ (z> \eta) $
the single or solitary bubble (Fig. 14a)
penetrates,
$ \eta $ is contact boundary
between liquid and "vacuum" bj$\infty. $
In Fig. 14a the right half of a bubble is drawn.
This bubble is symmetric
at transformation of a horizontal axis $x\to-x. $
Flow is symmetric
concerning a middle vertical straight line (b$\infty)$,
b is the apex of the bubble.
Penetration of a bubble
is accompanied by "squeezing" of liquid jets j.
On a vertical straight line (b$\infty)$
velocity
(arrows $w $ in Fig. 14b)
is positive
while
acceleration of Lagrangian particles
(arrows "acc", "dec")
changes a sign:
particles which are near to the bubble are decelerated,
and distant - are accelerated.

The profile of pressure with a maximum is shown in Fig. 14b.
We have put $p_b=p_{CB} =0$
Where $p_b $ is pressure in the apex b.
On infinity pressure in a liquid tends to the pressure in "vacuum"
$p_{CB}. $
The system of isobars 0, 1, 2, 3, 4 is represented in Fig. 14c.
The zero isobar 0 is the contact bj$\infty. $
Pressure grows with growth of number of an isobar.

\vspace{.5cm}
{\bf \# 5. Pressure and velocity distributions in a random case}

\vspace{.5cm}
Let's consider a random chain (Fig. 15).
Useful example of a random chain
(together with numerical calculation its correlation
and spectral characteristics)
is presented in work Inogamov et al., 2001b.
Horizontally homogeneous chains
(which are made from comparable bubbles)
are the most interesting.
Let $ \lambda_{ini} $ is the average bubble size.
The individual sizes of bubbles $ \lambda_i $ (Fig. 15)
are randomly scattered about average value $ \lambda_{ini} $
with a dispersion $ \Delta\lambda\sim \lambda_{ini}$.
Let's speak
"periodicity about randomness":
$Q =\lambda_{ini}/\Delta\lambda\sim 1$
where $Q $ is good quality (Q-factor).

Penetration of horizontally unbound bubble chains
raises pressure upon infinity $p_{\infty} $
in comparison with pressure upon boundary $p_{CB}. $
Pressure $p_{\infty} $ is created
by ensemble of bubbles
which is operating as the whole.
This pressure
is a measure of an average ram pressure $ \rho v^2 $
of bubbles in a chain.
Concerning squeezing of infinitely removed liquid
the action of the random chain is similar with periodic chain (\# 4).

In case of identical bubbles
(a periodic chain)
pressure in liquid is less than $p_{\infty} $
(Fig. 12, if $z<\infty$)
and heterogeneity of pressure is horizontally periodical.
The pressure created by solitary bubble
is more than $p_{\infty} $ (Fig. 14).
Movement of random ensemble
creates a pressure field $p(\vec r, t_{fix}) $
with a limit $p_{\infty}(t_{fix}) $
and random fluctuations
in which pressure can be both
or higher, or lower than limiting value $p_{\infty}$.

Important
that fluctuations $p$
form the hierarchical structure
penetrating deeply into bulk of liquid in space.
It is connected to formation of statistical groups of bubbles
with different size of a ram pressure, average for the given group.

{\bf $"\lambda$ groups" and $"\lambda$ fluctuations" of pressure.}
Let's describe
hierarchical structure of heterogeneity of an instant field of pressure.
Let's suppose
that the ensemble is placed in "box"
of the size $L\gg\lambda_{ini}$
with periodic boundary conditions at the box edges.
Let's break length $L $ into equal pieces
 $ \lambda\gg\lambda _ {ini} $
containing many bubbles $ ("\lambda$-group" of bubbles).
Groups of bubbles on pieces $ \lambda $ are various (Fig. 15).
Influence of bubble group
from a piece $ \lambda $
reaches a depth of the order $ \lambda $
from boundary into liquid.

In Fig. 16a
the figure 1
marks average position of contact boundary ("vacuum" is below).
The length of the pieces 3 is proportional to pressure
$p_{\lambda}$
in the center of a square $ \lambda\times\lambda $
Worth the bottom side on the boundary.
Changeability of this pressure
is caused by dissimilarity of $"\lambda$-groups" of bubbles.
The braces 2
show average pressure $ \langle p\rangle_x $
on distance $ \lambda/2 $ from the boundary.

Creation of momentum of liquid in square cells
$ \lambda\times\lambda $
is connected to fluctuations of pressure
$$
\Delta p_{\lambda} = p_{\lambda} - \langle p\rangle_x, \eqno (11)
$$
as the homogeneous component $ \langle p\rangle_x $ does not make movement.
Fluctuations of pressure (11) exponentially fade
at distances from boundary exceeding $ \lambda$.
Therefore at an estimation of momentum
it is necessary to take cells with height of the order $ \lambda. $

Let's write down an estimation of momentum
$$
\lambda^2 \rho w_{\lambda} \sim
(\rho w_{\lambda ini}^2) \lambda_{ini}
\sqrt{\lambda/\lambda_{ini}} \, t_{ini}. \eqno (12)
$$
In (12) $ \lambda^2$ is the area of a square,
and $w_{\lambda}$
is an increment of velocity of substance in this square
due to pressure action.
At the left side of equation (12)
there is a change of momentum in time $t_{ini}$
per unit length in a transverse direction.
On the right side
in parentheses the average ram pressure of bubble is written down.
Product of this pressure on $\lambda_{ini}$
gives average force per unit of transverse length.
With such force bubble presses on liquid.
Statistical multiplier
as a root from an average number of bubbles
in a $"\lambda$-group" $N_{\lambda}$
gives estimate of random fluctuations of pressure (11),
$N_{\lambda} = \lambda/\lambda_{ini}. $
The last multiplier on the right side of equation (12)
is time of existence of bubble generation of scale $\lambda_{ini}. $

As we see,
momentum bubble groups of scale $\lambda$
written down in the right part of balance (12)
diverges at large $\lambda$
because of the statistical factor $\propto \sqrt{\lambda}. $
It is caused by statistical accumulation of momentum
with growth of the size of $"\lambda$-groups".
Thus velocity of motion
$w_{\lambda} \propto I_{\lambda}/S_{\lambda}$ decreases,
here $I_{\lambda}$ and $S_{\lambda}$ are momentum
and the area of a $"\lambda$-group".

The sign of velocity fluctuations $w_{\lambda} $ (12)
randomly varies
from one square $\lambda\times\lambda$ to another (Fig. 16b).
Signs $w_{\lambda}$ and $\Delta p_{\lambda}$ (11) coincide.

{\bf Hierarchy of $"\lambda$-groups" and dependence from $z$.}
Above we have considered
one near boundary layer
of thickness $\sim\lambda$
in which the random velocity field $w_{\lambda}$
is created by $"\lambda$-groups" from $\approx N_{\lambda}$ bubbles.
Horizontal (random changeability)
and vertical (exponential fade) scales
of spatial heterogeneity of this field
is of order of the length of pieces $\lambda$, Figs. 15 and 16b.

To find velocity field from a random chain (Fig. 15)
it is necessary to consider set of splittings into pieces $\lambda$
of different scale.
The hierarchy of splittings
(or hierarchy of $"\lambda$-groups")
looks like: $ \lambda_0, \lambda_1, \lambda_2..., $
Where
$$
\lambda_n=a^n \lambda_{ini}, \eqno (13)
$$
$a> 1, a\sim 1, $ for example, $a=2. $
With each scale
(set by the index $n$)
particular near boundary layer is connected.
On the one hand
scales with different $n$ are independent
(there are not correlations between them),
on the other hand the final velocity field $w(z)$
(from a chain)
develops of imposing all layers $n.$

For an example we shall consider
$"\lambda$-groups" from ten (on the average) bubbles
$(N_{\lambda} = \lambda/\lambda_{ini} =10). $
Changeability on coordinate $x$ values $\Delta p_{\lambda} $ (11)
and $w_{\lambda} \propto \Delta p_{\lambda} $ (12)
is generated by non-equivalence of neighboring tens, Fig. 16a and b.
This changeability fades on depth $\sim 10\lambda_{ini}. $
According to the formula (12),
the amplitude of the given changeability makes
$w_{10} \propto 1/10^{3/2}. $

Let's compare $"\lambda$-groups" from tens and hundreds bubbles.
The amplitude of changeability of velocity
at shifts on coordinate $x$ on $100\lambda_{ini}$
is less $(w_{100} \propto 1/100^{3/2}). $
Therefore on depth $ \sim 10\lambda_{ini} $
fluctuations $w_{100} $
are imperceptible on a background of fluctuations $w_{10} $
(closer over contact is dominated with fineer splittings).
But on distances from boundary $ \sim 100\lambda_{ini}$
fluctuations $w_{10} $ exponentially decrease,
and to dominate begin fluctuations $w_{100}. $

Told it is illustrated in Fig. 17
on which the spatial map of vertical velocity $w$
on a plane of movement $(x, z)$ is shown.
Three consecutive scales $n-1, n, n+1 $
hierarchies (13) (small, average and big squares) are represented.
Velocities $w_{n-1}, w_n $ and $w_{n+1}$
dominate over different distances.
To negative velocities there correspond the shaded squares.
Thus, the field $w (x, z)$
looks like randomly branching tree
with a trunk $L,$
sharing on $a $ (13) branches
which in turn share for $a$ branches everyone, etc.

Dependence of amplitudes of fluctuations of velocity
on depth $w(z) $
is formed as result of "interference" of splittings $ \lambda_n. $
Total action of hierarchy of splittings (13) can be written down as
$$
w(z)=\sum_n w_n \exp \left( - \frac{2\pi z}{\lambda_n}\right) \propto
\sum_n \lambda_n^{-3/2}
\exp \left( - \frac{2\pi z}{\lambda_n} \right)
\propto \frac{1}{z^{3/2}},\;\;\;
w(z)\sim w_{\lambda ini} \frac{\lambda_{ini}^{3/2}}{z^{3/2}}. \eqno (14)
$$

Step-type behaviour
or lacunarity (13) (lacuna or gap) for hierarchy
was necessary for simplification of explanations.
It is possible
to consider an interference of splittings
with fine step of a geometrical progression
$a=1 +\epsilon, $ $ \epsilon\ll 1. $
Velocity field is smooth;
its particular shape
is defined by arrangement of bubbles in a chain (Fig. 15),
instead of hierarchy (13).
Thus of "stain" or "spots" of negative velocity $w$
have smooth contours (not squares, as in Fig. 17).
In Fig. 17
the diagram of splitting of consecutive scales
of these "spots", instead of their form is submitted.

The Fig. 18 supplements the formula (14).
Points $n-1, n $ and $n+1 $
mark the centers of squares, compare Figs. 17 and 18a.
These centers are on distances $ \lambda_n/2 $
from the boundary submitted by a horizontal piece.
The specified points are transferred on a plane $ (w, z) $
(Fig. 18b)
with the help of three horizontal arrows
connecting the left and right parts Fig. 18.
Coordinates of these points are
$ (w_n\propto 1/\lambda_n^{3/2}, z_n =\lambda_n/2). $
Through sequence of the transferred points
the curve $w(z) $ (14) is drawn.
As we see,
small-scale fluctuations $w_{\lambda} $ (12)
have the big amplitude
whereas large-scale will penetrate deeply into thickness of liquid.

Now it is possible to compare instant velocity fields $w (x, z) $
from a periodic chain (\# 4),
from a solitary bubble (\# 4),
and from a random chain (\# 5).
These
qualitatively various three cases
definitely supplement each other.
In the first case decay on $z $ occurs exponentially.
Power law dependence from $z$
is connected to a solitary (or lonely) bubble,
however distribution of velocity on coordinate $x $ is non-uniform.
Velocity field from a random chain,
first,
is homogeneous on transverse distances
large in comparison with the size of a bubble,
and, second,
as a power law depends on coordinate $z. $
As it will be visible below,
power law dependence $w $ from $z $
results in the power law of expansion
of a turbulent mixing zone
$(h\propto t^{\theta}). $
In a periodic case mixing quickly fades
(displacement of bubbles $h\propto \ln t). $

\vspace{.5cm}
\centerline{{\bf \# 6. Spectrum}}

\vspace{.5cm}
Let's calculate Fourier spectrum
of an instant velocity field $w (x, z=0) $ created by a random chain.
The scale
$$
 \lambda_{curr},\;\; k_{curr} =2\pi/\lambda_{curr},\;\;
 n_{curr} =L/\lambda_{curr}
$$
divides a spectrum on long-wave
 $ (n <n_{curr}) $
and short-wave $ (n> n_{curr}) $ parts.
The size $ \lambda_{curr} $
corresponds to scale of dominating structure
of the current generation of large-scale vorticies or the dipoles
developed by the moment of time $t\sim t_{curr},$
see Fig. 7, 8 and \# 3.
For questions examined in the given work
the short-wave part is unimportant.
The long-wave part
is determined by fluctuations $w_{\lambda} $ (12)
at $ \lambda> \lambda_{curr}. $

The scale of change of velocity $w_{\lambda} $
at length $ \lambda, $ see (6),
is connected to amplitudes of harmonics $w_n $ by equation (7).
Let's write down
$w_{\lambda} \sim \sqrt{n} w_n, $
where numbers of harmonics $n $ correspond to scale $ \lambda: $
$n\sim n_{\lambda} =L/\lambda. $
From here we receive
$$
w_n = w_{\lambda ini} \, (\lambda_{ini}/L)^{3/2} \,
\xi_n \cos\psi_n \, n, \eqno (15)
$$
where $ \xi_n\sim 1 $ is dimensionless number
and $ \psi_n $ is the random phase.
It turns out,
that a long-wave spectrum is linear under number $n. $
At $t\sim t_{curr} $ the formula (15)
is valid up to $n <n_{curr}$\footnote{
Instead of for $n <n_{ini} =L/\lambda_{ini}. $
In time $t\sim t_{curr} $
the maximum of a spectrum
is displaced from $n\sim n_{ini}$ in $n\sim n_{curr}. $
}.
At a derivation was considered,
that velocity field $w (x, z, t\sim t_{ini}) $
keeps the kind (under the order of size)
in scales $ \lambda> \lambda_{curr} $
down to the moments of time $t\sim t_{curr}$\footnote {
To discussion of this question are devoted \# 8 and \# 9 below.}

\vspace{.5cm}
\centerline{{\bf \# 7. Random chain and a shock}}

\vspace{.5cm}
We research velocity field $w^+(x, z)$
formed by a shock during short sound time $t_s $ (1)
near a random chain of linear perturbations $\eta^-$ (\# 1, \# 2).
Equation (5) connects harmonics of function $ \eta^+(x) $
and harmonics of initial velocity\footnote{
Initial for a stage of movement in a mode of an incompressible liquid.
}
Sequence of three steps of calculation following.
First, we define the Fourier spectrum $\eta^+_n$
of random surface perturbations $\eta^+(x)$\footnote{
It appears,
the long-wave part of a spectrum $ \eta_n $
does not depend from $n, $ $ \eta_n\propto n^0. $
}.
Second, under Richtmyer formula
the amplitude of harmonics $w_n$ is defined.
Thirdly, on a spectrum $w_n $ is function $w^+ (x, z)$ defined
and dependences of this function from $x$ and $z$ are analyzed.
Corresponding calculations are executed in work Inogamov, 2002.
It is shown, first,
similarly (15) spectrum $w_n $ is linear on $n,$
second, dependence $w $ from $z $ is given by the formula (14).

\vspace{.5cm}
{\bf \# 8. Decay of velocity with depth and expansion law}

\vspace{.5cm}
Let's address to a question on evolution of velocity field $w (x, z, t) $
(the given section and the following).
After a push connected to a shock wave,
there is velocity field $w^+(x, z) $ (\# 7).
We admit, in the further field $w^+$ remains "frozen" in liquid.
In the following section we shall return to this assumption.
Let's explain, that means "frozen".
Basis is that
(as it was spoken \# 5)
small-scale modes have large velocities,
and large-scale will penetrate deeply into liquid.

The field of velocity $w^+(x, z)$
with enlarging cells
(Figs. 17, 18a)
and the power law decay on $z$ (Fig. 18b) is formed.

At such position the assumption about "frozen" cells is natural.
Really,
turning time of large formations is large
$$
t_{\lambda} \sim \lambda/w_{\lambda} \propto t^{5/2}. \eqno (16)
$$
Change of velocity $w_{\lambda}$
on depth $z\sim\lambda$
occurs {\it after penetration of contact boundary
on this depth $h_+\sim\lambda.$}
If it so it is possible to write down
$$
\frac{dh_+}{dt} \sim w(z)|_{z\sim h_+} \sim 1/h_+^{3/2}.
$$
From here follows the required law\footnote{
Several phenomenological approaches
was earlier offered
(K-model and buoyancy-drag model)
for the approximate estimation of an exponent $ \theta.$
Usually
reduction of this exponent
in comparison with value 2/3 (conservation of energy)
is connected with viscous losses of energy on small scales.
The review of these approaches is given in work Inogamov 2002.

In work of Schwarz et al., 2000
the exponent $\theta$ for 2D and 3D cases
has been extimated with the help of bubble envelop model.
The envelop model
essentially distinguishes
from K-model and buoyancy-drag model.
Values $\theta_{2D} \approx 0.4$
and    $\theta_{3D} \approx 0.2$
have been received by the envelop model (Schwarz et al., 2000).

As against the listed approaches
in the present work
the exponents $\theta_{2D}$ (17)
and $\theta_{3D}$ (\# 12 below)
are calculated precisely.
The mechanism of expansion of mixing zone is opened.

Appeared,
dynamics of expansion
is defined by long wavelength $ (\lambda> \lambda_{curr})$
fluctuations of velocity spectrum.
At the moment of time $t_{curr}$
these fluctuations
are at a linear stage of development
(displacement of liquid particles $\eta_{\lambda}$
on scale $\lambda$
are small in comparison with $\lambda).$
}
of decay of Richtmyer-Meshkov turbulence
$$
h_+\sim w_{\lambda ini}^{2/5} \,\lambda_{ini}^{3/5}\, t^{\theta},
 \; \; \; \; \;
\theta = 2/5. \eqno (17)
$$
Large-scale formations
enter game one by one
in the hierarchical order (13)
and with the big delay (16).
This gradual inclusion of more and more large cells
defines the law $h_+(t)$ (17).

\vspace{.5cm}
{\bf \# 9. Accumulation of momentum and logarithmic corrections}

\vspace{.5cm}
Velocity field $w^+(x, z) $
is formed (\# 7)
for a time interval $\sim t_s $ (1).
Further in time
$\sim t_{\lambda ini}$ (2)
"balls" (Fig. 8) primary vorticies
are turned off -
the dominating structure of the first generation is formed.
At $t\sim t_s$ displacement of contact boundary
are small (still there is no turning boundary).
The random chain of sign-variable vorticies
$\lambda_{ini}, w_{\lambda ini}$
forms velocity field $w(x, z, t\sim t_{\lambda ini})$
in near boundary layer.
Important that fields $w^+(x, z)$ and $w(x, z, t_{\lambda ini})$
under the order of size are identical.

Further process of change of one generations by the following develops.
Let's consider the current generation
with scale of dominating structure $\lambda_{curr}.$
The random chain $\lambda_{curr}, w_{\lambda curr}$
forms velocity field
$$
w(x, z, t\sim t_{\lambda curr}) \sim
w_{\lambda}(t_{\lambda curr}) \sim
w_{\lambda curr} (\lambda_{curr}/\lambda)^{3/2} \eqno (18)
$$
in surpassing scales $ \lambda> \lambda_{curr}.$
Substituting in (18)
following from the law (17) estimations
$$
\lambda_{curr} \sim h_+(t_{\lambda curr}),
 \; \; \; 
w_{\lambda curr} \sim
 \frac{dh_+(t_{\lambda curr})}{dt},
$$
we express velocities $w(x, z, t\sim t_{\lambda curr})$
or $w_{\lambda}(t_{\lambda curr})$ in (18)
through parameters of primary vorticies $\lambda_{ini}, w_{\lambda ini}.$
It turns out,
that in the rest by the moment $t\sim t_{\lambda curr}$
undeveloped (or "unsaturated")\footnote{
The latent or "frozen" scales.
}
scales $\lambda> \lambda_{curr}$
all generations of dominating structures
create under the order of size the same field of velocity $\sim w^+!$

At discussion of velocities $w(x, z, t_{\lambda curr})$
or $w_{\lambda}(t_{\lambda curr})$
the question was amplitudes of fluctuations.
Let's consider now the phase information.
On amplitude velocities
created by different generations
are of the same order of magnitude value.
It is asked,
whether the fields
created by different generations
are identical on a phase?
In other words,
whether the shaded cells on a map of velocities
from different generations (Fig. 17) coincide?

There are three opportunities.
First,
the large-scale part $ (\lambda> \lambda_{curr})$
fields $w(x, z) $
can practically not vary in time.
In this case the law of mixing (17) is carried out.
Second,
structure of shaded cells
formed by different generations
can coincide under the order of magnitude.
Thirdly,
this structure can randomly vary
from generation to generation.

Let's consider last two opportunities.
Let's recollect,
that $w_{\lambda}(t_{\lambda curr}) $
is {\it an increment} of velocity
connected to an increment of momentum
for a time interval $\sim t_{curr}. $
Means for the finding of velocity field
developing by the moment of time $t\sim t_{curr}$
it is necessary to summarize all increments
caused by activity of the previous generations.

In case the second opportunity is realized, we receive
$$
w_{\lambda}(t_{\lambda curr}) \sim
\ln (\lambda_{curr}/\lambda_{ini}) \, w^+. \eqno (19)
$$
In the third case we have
$$
w_{\lambda}(t_{\lambda curr}) \sim
\sqrt {\ln (\lambda_{curr}/\lambda_{ini})} \, w^+. \eqno (20)
$$
Accumulation of increments
results in logarithmic amplification.
The logarithmic corrections
are connected with a little bit strengthening of mixing.
In view of it instead of the law (17) it is received
$$
h_+\propto (\ln t)^{2/5} \; t^{2/5}, \; \; \; \; \;
h_+\propto (\ln t)^{1/5} \; t^{5/2}
$$
in cases (19) and (20) accordingly.

In 3D case we have
$$
h_+\propto (\ln t)^{1/3} \; t^{1/3}, \; \; \; \; \;
h_+\propto (\ln t)^{1/6} \; t^{1/3}.
$$

\vspace{.5cm}
\centerline{{\bf \# 10. Influence of inertia of light phase}}

\vspace{.5cm}
In the previous sections
the case $\mu=0$
(At $=1,$ contact to "vacuum", $\mu$ is density ratio)
was investigated.
The case $ \mu\neq 0 $
(the account of density of light phase)
is difficult, and demands separate consideration.
Position at $ \mu\neq 0$ in 2D geometry is difficult.
In this case kinetic energy of an incompressible liquid conserves\footnote{
Let's note, that in a case $\mu=0$ viscous dissipation
at Re$\to\infty $ is absent in 3D geometry.
}
for the case of small viscosity (Re $\gg 1).$
In a case $\mu=0$
the viscous heat production can be neglected
at large values of Reynolds number Re
(conservation of kinetic energy).
However
difficulty with excess of energy in small scales\footnote{
Formally
the law of conservation of energy
results in value $ \theta=2/3$
which essentially exceeds the value (17).
}
at $\mu=0$ does not arise.
The matter is that
the fly of fast jets in "vacuum" carries away surplus of energy.
At such position
promotion of front $h_+ $
aside heavy liquid
is defined
by the law of conservation of momentum
which results in the formula (17).

Flow with the power law of expansion $h_+\propto t^{\theta}$
is self-similar.
Thus in self-similar variables
$$
t^{\theta}\, n,\; \; \, t^{1-3\theta/2}\, w_n 
$$
spectrum (\# 6) is steady-state\footnote{
It is steady-state on the average.
Thus there are fluctuations of amplitudes
about average values and fluctuations of phases.}.
The degree scaling amplitudes,
appears from an equation:
$ \sqrt{n}\, w_n\sim
 \dot h_+$ at $n\sim n_{curr},$ see (7)
and \# 6 (definition of $n_{curr}). $

In self-similar flow
the profile of average density
$\bar\rho = \langle\rho\rangle_{\bot}$
is function not of two $(z, t)$
but of one variable\footnote{
Certainly,
self-similar asymptotics
is established in process of growth of number of the generations
replaced
after generation of primary vorticies
$\lambda_{ini}, w_{\lambda ini}.$
},
$\bar \rho(z/h_+)$
or $\bar \rho (z/t^{\theta});$
the mark $\bot$ means averaging on transverse coordinates.
The coordinate $z$
is counted from initial position of boundary (\# 1, \# 2).
At $ \mu=0 $
function $ \bar\rho $ quickly decreases aside "vacuum".
It becomes very small
already at value of dimensionless coordinate
$(-z)/h_+(t)$
about several units.

First portions of the substance
(which have been thrown out by primary generation of bubbles)
flying ahead aside "vacuum"
move under the law $h_-\sim - w_{\lambda ini} t.$
Distribution of density near to front $h_-$
is not described by the self-similar solution $ \bar \rho (z/h_+).$

\vspace{.5cm}
\centerline{{\bf \# 11. Stimulated and spontaneous turbulence}}

\vspace{.5cm}
{\bf Addition of initial long wavelength spectrum.}
Problem I is above considered
in which the long wavelength part
$(\lambda> \lambda_{curr})$
of a spectrum $w_n $ or $w_{\lambda}$
is formed owing to an interference of random small-scale modes.
Let's address to other problem
(a problem II)\footnote{
Below problems I and II refer to
as problems about stimulated and spontaneous turbulence.
}.
In it
to a random initial chain of surface perturbations
$\lambda_{ini}, \eta_{\lambda ini}$
the long wavelength spectrum is added
$$
\eta^{LW}_n =
\frac{
\lambda_{\star}^{\beta+3/2}
}
{L^{\beta+1/2}}\,
\xi_n\, \cos\psi_n\; n^{\beta},\;\;\;\;\;
1\leq n\leq n_{ini}=L/\lambda_{ini},
\eqno (21)
$$
where $\langle\xi_n\rangle =1,$
$\psi_n $ is a random phase,
$L $ is length of "box" (\# 2).
Let's write down also
$$
\eta^{LW}_{\lambda}=
(\lambda_{\star}^{\beta+3/2}/L^{\beta+1/2})\, n^{\beta+1/2},\;\;
\epsilon^{LW}_{\lambda}=\eta^{LW}_{\lambda}/\lambda\sim
(\lambda_{\star}/L)^{\beta+3/2}\, n^{\beta+3/2}
\sim (\lambda_{\star}/\lambda)^{\beta+3/2},
\eqno (22)
$$
where $\epsilon^{LW}_{\lambda}$
is a relative amplitude of surface perturbation.
It characterizes an angle
of a deviation of a tangent to perturbed surface
from unperturbed position.

At $\lambda\sim \lambda_{\star}$
we have $ \eta^{LW}_{\lambda} \sim \lambda $ (22) -
the assumption of linearity of perturbations is broken.
As we are interested with long wavelengths
$ \lambda> \lambda_{\star},$
should be
$$
\beta>-3/2. \eqno (23)
$$
Value
$$
\beta = - 3/2 \eqno (24)
$$
corresponds to a {\it homogeneous} spectrum of perturbations
$$
\eta^{LW}_n\sim \epsilon_{hom} L/n^{3/2}\propto 1/|k_{\bot}|^{3/2},\;\;\;\;
\eta^{LW}_{\lambda}\propto 1/n,\;\;\;\;
\epsilon^{LW}_{\lambda}\sim \epsilon_{hom}={\rm const,} \eqno (25)
$$
where $k_{\bot}$ is transverse or horizontal wave number.
Homogeneous noise (25)
are dimensionless in the sense
that do not contain parameter $ \lambda_{\star} $
with dimension of length
as in this case it is carried out
$\eta^{LW}_{\lambda} \sim \lambda.$
They are homogeneous in space of wave numbers
as have no the allocated scale (from here the name).
These noise
play the important role in the theory of the turbulence
connected to Rayleigh-Taylor instability.

Let's "sew"
a long wavelength part of the initial spectrum
set by formulas (21), (22)
with perturbations
"located" at wave number axis
$\lambda_{ini}, w_{\lambda ini}$\footnote{
The short wavelength tail
$ \lambda <\lambda_{ini}$
is not significant for dynamics of system at $t\gg t_{\lambda ini}. $
The long wavelength tail
created by "located" at $k\sim k_{ini}$ perturbation (Fig. 15)
looks like $w_n\propto n $ (15).

Long wavelength spectra (15) and (21) are different.
}.
Then amplitudes $w^{LW}_{\lambda}$
(curves 1 and 3 in Fig. 19)
and $w_{\lambda ini}$
(a curve 4)
are of the same order
at $\lambda\sim\lambda_{ini}.$
Let's write out a condition of sewing together
$\epsilon^{LW}_{\lambda} \sim \eta_{\lambda ini}/\lambda_{ini}$
having taken $\epsilon^{LW}_{\lambda}$ in this condition
at $ \lambda\sim \lambda_{ini}.$
Let's express parameter $ \lambda_{\star} $
through $ \lambda_{ini} $ with the help of the given condition
$$
\lambda_{\star} \sim
(\eta_{\lambda ini}/\lambda_{ini})^{1/(\beta+3/2)} \, \lambda_{ini}.
$$
We see,
at $ \eta_{\lambda ini} <\lambda_{ini} $
for the spectra limited to a condition (23)
is carried out $ \lambda_{\star} <\lambda_{ini}, $ see Fig. 19.

Excluding $ \lambda_{\star}$ it is received
$$
\eta^{LW}_n=\epsilon_{ini} \frac{
\lambda_{ini}^{\beta+3/2}
}
{L^{\beta+1/2}}\,
n^{\beta},        \;\;
\eta^{LW}_{\lambda}=
\epsilon_{ini} \frac{
\lambda_{ini}^{\beta+3/2}
}
{L^{\beta+1/2}}\,
n^{\beta+1/2},\;\;
\epsilon^{LW}_{\lambda}=
\epsilon_{ini} \left(\frac{
\lambda_{ini}
}
{L}\right)^{\beta+3/2}\,
n^{\beta+3/2}, \eqno (26)
$$
Where the notation
$\epsilon_{ini} = \eta_{\lambda ini}/\lambda_{ini}$
is entered.
Equations (26)
describe added\footnote{
Added in the initial data
to "located" on wave number $k$ initial perturbation.
This is the combined
("located" and long wavelength)
or compound perturbation.
}
long wavelength initial "noise".
They replace expressions (21) and (22).

{\bf Development of Richtmyer-Meshkov turbulence
at presence of added long wavelength "noise".}
In Fig. 19 perturbation $\lambda_{ini}, w_{\lambda ini}$
(a curve 4)
"located" on an axis $n $
is shown.
Large-scale fluctuations of the "located" perturbation
generate (\#\# 5-9) a long wavelength part of a spectrum.
It is represented by a curve 2.
Thus, the spectrum representing association of curves 2 and 4
is connected to the "located" perturbation.

In Fig. 19
added long wavelength "noise"
is represented with curves 1 and 3
concerning to different values of an index $\beta$ (26).
For an example from a curve 1
we have $\beta <0, $
and for a curve 3 it is carried out $ \beta> 0.$
At $ \beta=0 $
dependences $ \epsilon_{\lambda} $ from $\lambda $
are identical to added long wavelength "noise" (a curve 2)
and for large-scale fluctuations of the "located" perturbation (a curve 2).
Let's note,
that under condition of (23)
for these dependences the limiting equation
$ \epsilon_{\lambda} \to 0$
is carried out
at $ \lambda\to\infty. $

Interest represents a case
$-3/2\leq\beta <0$
when added long wavelength "noise"
surpasses large-scale fluctuations of the "located" perturbation.
Such "strong" noise defines dynamics of mixing
at $t\gg t_{\lambda ini}. $

The harmonics of velocity
corresponding to this "noise"
are defined by Richtmyer formula (5)
$$
w^{LW}_n\sim (2\pi/L) n \eta^{LW}_n W_{unpertCB} \propto n^{1 +\beta}.
$$
For large-scale fluctuations of velocity it is received
$$
w_{\lambda} \sim \sqrt{n} w_n\propto n^{\beta+3/2}. \eqno (27)
$$
Let's note,
that according to Richtmyer formula
it is carried out $w_{\lambda} \propto\epsilon_{\lambda}. $

Let's write down
$dh_+/dt \sim w_{\lambda curr}$
(see \# 8 and \# 9).
We have
$h_+\sim \lambda_{curr} $ (Figs. 6-8).
From here we receive
$$
h_+\sim \lambda_{curr}\sim \epsilon_{ini}^{1/(\beta+5/2)}\,
\lambda_{ini}^{(\beta+3/2)/(\beta+5/2)}\,
W_{unpertCB}^{1/(\beta+5/2)}\,
t^{1/(\beta+5/2)}
\sim 
$$$$
\sim \lambda_{ini}^{(\beta+3/2)/(\beta+5/2)}\,
w_{\lambda ini}^{1/(\beta+5/2)}\,
t^{1/(\beta+5/2)},
\eqno (28)
$$
as
$w_{\lambda curr} \propto 1/\lambda_{curr}^{\beta+3/2}$
according to (27).

Homogeneous noise (24), (25) is
the extreme case of the large-scale noise
stimulating mixing.
In this case
during the initial moment
we have $w_{\lambda} \sim w_{\lambda ini}$
at anyone $ \lambda> \lambda_{ini}. $
Hence, the formula (28) becomes $h_+\sim w_{\lambda ini} t. $

{\bf Comparison of stimulated and spontaneous mixing.}
There is a certain analogy
between classes of turbulent movements
at Rayleigh-Taylor and Richtmyer-Meshkov instabilities.
There is a division on
stimulated and spontaneous classes
in Rayleigh-Taylor case
(Inogamov, Demianov, and Son, 1999; Inogamov, 1999)
and analogue of this division
in Richtmyer-Meshkov case\footnote{
According to it
let's carry problem I (see the beginning \# 11)
to spontaneous Richtmyer-Meshkov turbulence
and a problem II - to stimulated.
}.
Let's remind sense of concepts
about stimulated and spontaneous motions.
At spontaneous turbulence
unstable boundary
is initially "seeded"
with the perturbations
"located" on $k $ or $ \lambda $ near $ \lambda_{ini}. $
Are interested in an self-similar regime
which is established asymptotically
at $h_+\gg \lambda_{ini}. $
Flow tends to a self-similar expansion regime
of a mixing zone
$$
h_{\pm} = \alpha_{\pm} {\rm At} g t^2 \eqno (29)
$$
with factors $ \alpha_{\pm}^{sp}$
not dependent on a concrete shape of "located" on $k$ initial perturbations
("forgeting" of initial perturbations $ \sim\lambda_{ini}).$

We call stimulated a regime with long wavelength stimulation.
Imposing at $t=0 $
long wavelength power law spectra
such as (26)
introduces scale $\lambda_{\star}$
and thus "breaks" self-similar dependence (29).
The special place belongs to homogeneous "noise"
with perturbations of boundary
$ \eta $ a kind (25)
and/or velocity perturbations of a kind
$$
w_n\sim \epsilon_{hom}^v \sqrt{{\rm At}gL}/n\propto 1/|k_{\bot}|,\;\;\;
w_{\lambda}\sim \epsilon_{hom}^v \sqrt{{\rm At}g\lambda}. \eqno (30)
$$
In 3D a case
(see also \# 12 below)
we have the following expressions for homogeneous noise
$$
\eta_{nm}\sim \frac{\epsilon_{hom} L}{N_{nm}^2}\propto |k_{\bot}|^{-2},\;\;\;
\eta_{\lambda}\sim \frac{\epsilon_{hom} L}{N_{nm}},\;\;\;
w_{nm}\sim \epsilon_{hom}^v \frac{\sqrt{{\rm At}gL}}{N_{nm}^{3/2}}\propto
 |k_{\bot}|^{-3/2}.
 \eqno (8.30)'
$$
Homogeneous noise do not contain dimensional parameter
$ \lambda_{\star}. $
Hence, self-similarity (29) it is kept.
However,
with "noise" (25), (30), (30)'
dimensionless numbers or amplitudes are connected
$\epsilon_{hom}, \epsilon_{hom}^v. $
In such situation
factors $ \alpha_{\pm} $
become functions
$ \alpha_{\pm}^{st}(\epsilon_{hom}, \epsilon_{hom} ^v) $
of these dimensionless amplitudes
describing intensity of "noise" stimulating mixing.

Real perturbation
is a combination of spontaneous and stimulated\footnote{
In cases with small values of stabilizing factors
(viscosity, heat conductivity, diffusion)
spontaneous turbulence gets ineradicable character
while stimulated turbulence -
it is supervised by dimensionless amplitudes $ \epsilon_{hom}. $
}
perturbations.
Dependence
$ \alpha_{\pm}^{st}(\epsilon_{hom}, \epsilon_{hom}^v) $
is slowly growing function of the arguments
(a curve "st" in Fig. 20).
The factor
$ \alpha_+^{sp+st}(\epsilon_{hom})$
(a curve "sp+st")
is limited from bottom\footnote{
Due to reduction of $\epsilon_{hom}$
it is impossible to achieve that
mixing will develope more slowly than spontaneous mixing
(the spontaneous regime "props up" the combined modes "from below").
Similar situation exists
for the Richtmyer-Meshkov mixing, see Fig. 21 below.}
by the constant value
$ \alpha_+^{sp} $ (a straight line "sp").
From the data on factor $ \alpha_+^{sp}$
suggested in report Youngs 2004
in Fig. 20 value of $0.03$ is accepted.
There is a value of dimensionless or relative amplitude
$ \epsilon_{hom}^{thr}$
at which influence of stimulation becomes appreciable.

It is very important
that numerically the value
$ \epsilon_{hom}^{thr} $ is rather small.
It was supposed in the previous estimations
(Inogamov, Demianov, and Son, 1999; Inogamov, 1999)
and has been confirmed in recent detailed numerical calculations
(Youngs 2004).
Smallness of the factor $\alpha_+^{sp}$
and smallness of the threshold amplitude
$\epsilon_{hom}^{thr}$
strongly raise a role of a regime with stimulation.
The amplitude $ \epsilon_{hom} $
gives value of deviations
of a tangent to boundary on all scales
as at homogeneous noise $ \epsilon_{\lambda} $
does not depend from $ \lambda. $
Smallness of $\epsilon_{hom}^{thr}$ means
that hidden or latent widezone perturbations are significant.
It is caused by that
these weak widezone noise can increase
essentially
the major characteristic of mixing - the factor $ \alpha_+. $

In reports Youngs 2004 and Dimonte et al. 2004
it is assumed,
that available divergences
between numerical $ (\alpha_+\approx 0.03) $
and experimental $ (\alpha_+\approx 0.05) $
estimations $ \alpha_+ $
are caused by the latent stimulation of mixing in experiment.
Such stimulation difficultly to supervise
because of smallness of $ \epsilon_{hom}^{thr}. $
The origin of $ \epsilon_{hom} $
can be connected to vibration of experimental installations.
The interesting program
(Dimonte et al., 2004)
for search of weak widezone noise in experiment is offered.

{\bf Analogue of stimulation of mixing in Richtmyer-Meshkov case.}
At Richtmyer-Meshkov turbulence
large-scale perturbations (26) intensify mixing.
In stimulated regime
(fig. 21)
the exponent $ \theta $ is more,
than in a spontaneous regime:
$ \theta^{st}> \theta^{sp}, $
$ \theta^{st} = (\beta+5/2)^{-1}, $
see (28) \footnote{
In report Youngs 2004
stimulation of Richtmyer-Meshkov mixing is numerically investigated.
Power law initial surface perturbations of a kind (26) has been used.
Change of the exponent $\theta$
is revealed at change of an index $\beta. $
},
$\theta^{sp} =2/5, $ see (17), \# 8, \# 9.
Borders of an interval of values of an index $ \beta $
form points $ \beta=0 $ and $ \beta = - 3/2 $ on an axis $ \beta. $
In Fig. 21 more to the right of the specified interval
the forbidden area, see (23) settles down.
More to the left - the area lays,
in which stimulation it is possible to neglect
(compare spectra 2 and 3 in Fig. 19).
The point $ \beta = - 3/2 $ (24)
represents homogeneous or "dimensionless" "noise" (25).
To them there corresponds
the maximal value of function $ \theta^{st}(\beta.) $

The most important\footnote{
As this point represents the most widespread variant of initial conditions
With "localization" of perturbations
on wave number $k $ (a random chain $ \lambda_{ini}). $
}
is the point $ \beta=0$.
It
concerns to a mode
which we shall conditionally name "spontaneous".
The reserve here contains in the following.
The comparative analysis
of Rayleigh-Taylor and Richtmyer-Meshkov problems
is carried out.
At Rayleigh-Taylor turbulence
in a spontaneous case
there is no dynamically essential long-wave tail
existing on "all times" $t> 0 $
(to it is connected "forgeting"
small-scale initial perturbations).
On the contrary,
in the stimulated case
such tail is present
(constant presence of a part $ \lambda> \lambda _ {curr} $
of an initial spectrum).
In this respect
at Richtmyer-Meshkov mixing business is in another way.
Namely,
both in stimulated and in spontaneous cases
there is constantly existing long wavelength tail
determining dynamics.
Hence, and in a spontaneous case
the certain "trace" of the initial data is kept.

Richtmyer-Meshkov turbulence
in a regime (17)
is generated by random "located" on $k$ perturbation
(\#\# 5-8).
Such perturbation
generates spontaneous mixing in Rayleigh-Taylor case.
The basis is connected to it
to name spontaneous a regime of mixing (17) in Richtmyer-Meshkov case.

\vspace{.5cm}
\centerline{{\bf \# 12. 3D case}}

\vspace{.5cm}
Let's generalize the description of turbulent mixing
received above on 3D space.
The horizontal plane with coordinates $x, y $
(Fig. 2) is transverse plane.
Let's limit a part of this plane
by big
$ (L\gg\lambda_{ini}) $
square $L\times L $ ("box", \# 2).
Let's consider
a random two-dimensional lattice of perturbations
on a plane $x, y. $
Such lattice (3D)
replaces a random chain (2D, \# 5).
Let the average transversal size
of initial surface perturbations
$$
z = \eta_{ini}(x, y) \eqno (31)
$$
in this lattice
is equal to $\lambda_{ini}$
and randomness is characterized by good quality $Q\sim 1$
(\# 5).

Let's expand
periodic in a square $L\times L $ function (31)
in Fourier series
$$
\eta_{ini}(x,y)=\sum_{n=1}^{\infty}\sum_{m=1}^{\infty} (\eta_{nm}C_{nx}C_{my}+
\eta_{nm}'C_{nx}S_{my}+\eta_{nm}''S_{nx}C_{my}+\eta_{nm}'''S_{nx}S_{my}),
 \eqno (32)
$$
Where $C_{nx} \equiv \cos (2\pi n x/L), $
$S_{nx} \equiv \sin (2\pi n x/L). $
Amplitudes of Fourier harmonics
in series (32) are calculated by integration
$$
\eta_{nm}=(4/L^2)\int_0^L\int_0^L dx \,dy\,\eta_{ini}(x,y)C_{nx}C_{my}. \eqno (33)
$$
Similar formulas
define other amplitudes which are included in a series (32).

Long wavelength $ (n, m\ll N_{ini} =L/\lambda_{ini})$
asymptotics of spectra $ \eta_{nm} $ looks like
$$
\eta_{nm}\sim \xi_{nm} \cos\psi_{nm}
(4/L^2) \eta_{rms} \lambda_{ini}^2
\sqrt{\lambda^2/\lambda_{ini}^2}
\sqrt{L^2/\lambda^2}\sim \eta_{rms}\lambda_{ini}/L\sim \eta_{rms}/N_{ini},
\eqno (34)
$$
where $ \langle\xi_{nm} \rangle=1, $ $ \psi_{nm} $ is a random phase,
$ \eta_{rms}^2 = (1/L^2) \int_0^L\int_0^L dx \, dy \,\eta_{ini}^2, $
$ \lambda=2\pi/k, $ $k^2 = (k^x_n)^2 + (k^y_m)^2, $
$k^x_n=2\pi n/L, $ $k^y_m=2\pi m/L $ are components of a wave vector
on a plane $ (x, y). $
This vector corresponds to a harmonic $ (n, m). $
Asymptotics (34)
follows from a statistical estimation of integral (33)
of stochastic function (31).
Occurrence of statistical factors
is connected to stochastic fluctuations of this function
$\sqrt{\lambda^2/\lambda_{ini}^2} $ and $ \sqrt{L^2/\lambda^2}. $
The estimation (34) is valid on an interval of values $n, m $
from 1 up to $ \sim N_{ini}. $

In 3D a case instead of (7) we have
$\eta_{\lambda}\sim \sqrt{n^2+m^2}\eta_{nm}\sim \eta_{rms}\lambda_{ini}/\lambda,$
where $n\sim m\sim L/\lambda. $

Let perturbation (31) is linear (small amplitude):
$ \eta_{rms} \ll \lambda_{ini}. $
Let's estimate velocity field after passage of a shock wave.
We use Richtmyer formula
connecting amplitudes of harmonics of surface perturbations
$ \eta_{nm} $ and vertical velocity $w_{nm}. $
In 3D geometry we have
$$
w_{nm}=F {\rm At}^+ (2\pi/L) \sqrt{n^2+m^2} \eta^+_{nm} W_{unpertCB}, \eqno (8.35)
$$
compare with the formula (5).

Let's find velocity of movement of primary bubbles $w_{\lambda ini}. $
Such bubbles have transverse scale $ \sim\lambda_{ini}. $
Let's substitute in the formula (35)
expression for amplitude $ \eta_{nm} $ (34) at $n\sim m\sim N_{ini}. $
In result it is received
$$
w_{nm}(n,m\sim N_{ini})\sim F {\rm At}^+ (\eta_{rms}/L) W_{unpertCB}. \eqno (36)
$$

Let's write down
$$
w_{rms}\sim w_{\lambda ini}\sim
 w_{\lambda}(\lambda\sim\lambda_{ini})\sim N_{ini} w_{nm}(n,m\sim N_{ini}) \eqno (37)
$$
and
$$
w_{\lambda}(\lambda\sim\lambda_{nm}\gg\lambda_{ini})\sim (L/\lambda_{nm}) w_{nm},\;\;\;\;
\lambda_{nm}=L/\sqrt{n^2+m^2}. \eqno (38)
$$
In the formula (38)
size $w_{\lambda} $
is determined by the formula (6) generalized on 3D a case.
Formulas (37), (38) replace the formula (7) in 3D a case.
Fourier amplitudes $w_n $ (2D) or $w_{nm} $ (3D)
are connected to the corresponding size $w_{\lambda} $
by means of the statistical factor
of order of a root square from number of modes of scale $ \lambda. $
In 2D case this number is $L/\lambda$
and in 3D a case - is $ (L/\lambda)^2. $

From formulas (36), (37)
it is possible to define velocity of primary bubbles $w_{\lambda ini}. $
The estimation of this velocity looks like
$$
w_{\lambda ini}\sim F {\rm At}^+ N_{ini} (\eta_{rms}/L) W_{unpertCB}. \eqno (39)
$$

Asymptotics of 3D Richtmyer-Meshkov mixing
at late times is defined by a long wavelength part
$ (n, m\ll N_{ini}) $ of a spectrum of velocity $w_{nm}. $
Let's find asymptotics of a spectrum $w_{nm} $
with the help of asymptotics of a spectrum $ \eta_{nm} $ (34),
Richtmyer formula (35) and auxiliary expression (39).
We have
$$
w_{nm}\sim (\sqrt{n^2+m^2}/N_{ini}^2) w_{\lambda ini}. \eqno (40)
$$
It is useful to compare 3D expression (40) to the formula (15)
valid in 2D case.

From formulas (38) and (40)
the asymptotics $(\lambda\gg \lambda_{ini}) $ of velocity fluctuations
of scale $ \lambda $
can be found.
It may be written as
$$
w_{\lambda}(\lambda\sim\lambda_{nm}\gg\lambda_{ini})\sim \frac{n^2+m^2}{N_{ini}^2} w_{\lambda ini}\sim
\left(\frac{\lambda_{ini}}{\lambda}\right)^2 w_{\lambda ini}. \eqno (41)
$$

The hierarchy (13) of fluctuations (41)
of different scales
forms enlarging cell structure
of velocity field $w(x,y,z)$
similar to presented in Fig. 17.
Three-dimensional cells
have the sizes $ \sim\lambda\times\lambda\times\lambda. $
As well as in 2D a case,
decrease of amplitude of fluctuations of 3D fields $w(x, y, z) $
at removal from boundary
is described by the power law.
To tell the truth, an exponent is another.
From the formula (41) it is received
$$
w(z)\sim w_{\lambda ini} (\lambda_{ini}/z)^2, \eqno (42)
$$
compare with (14).

Decay of velocity $w(z) $
defines the law of penetration of mixing front $h_+(t)$.
Let's write down $dh_+/dt \sim w(z\sim h_+)$ (see \# 8).
Substituting here the formula (42) and integrating
we receive
$$
h_+\sim w_{\lambda ini}^{1/3} \lambda_{ini}^{2/3} t^{\theta},
\;\;\;\;\theta=1/3. \eqno (43)
$$
Let's compare laws (43) and (17)
concerning to 2D and 3D to cases.
We see, in a three-dimensional case
Richtmyer-Meshkov turbulence fades faster
and the front $h_+(t) $ moves ahead essentially more slowly
$$
\theta _ {3D} <\theta _ {2D},
$$
where values $ \theta_{2D} $ and $ \theta_{3D} $
are given by formulas (17) and (43) accordingly.

\newpage
\centerline{{\bf References}}

\noindent M. Chertkov, 2003,
Phenomenology of Rayleigh-Taylor turbulence,
Phys. Rev. Lett., vol. 91(11), 115001 (2003).

\noindent V. Demchenko, 2004,
Colliding surface instability for a high velocity impact,
9th International Workshop on the Physics of Compressible Turbulent Mixing,
Cambridge, UK, 19-23 July 2004, Abstracts, p. 24;
http://www.damtp.cam.ac.uk/iwpctm9/

\noindent G. Dimonte, P. Ramaprabhu, and M. Andrews, 2004,
Dependence of self-similar Rayleigh-Taylor growth on initial conditions,
9th International Workshop on the Physics of Compressible Turbulent Mixing,
Cambridge, UK, 19-23 July 2004, Abstracts, p. 26;
http://www.damtp.cam.ac.uk/iwpctm9/

\noindent N.A. Inogamov, 1999,
The Role of Rayleigh-Taylor and Richtmyer-Meshkiv Instabilities in Astrophysics: An Introduction,
Astrophysics and Space Physics Reviews, vol. 10, part 2, 1-335 (1999).

\noindent N.A. Inogamov, A.Yu. Demianov, and E.E. Son, 1999,
Hydrodynamics of mixing:
Periodical structures,
Amplification of subharmonics
and inverse cascad,
in Russian,
Moscow, MIPT, 1999, 464 p.

\noindent N.A. Inogamov and A.M. Oparin, 1999,
Three-dimensional array structures associated with
RichtmyerЦMeshkov and RayleighЦTaylor instability,
Journal of Experimental and Theoretical Physics, vol. 89(3), 481-499 (1999).

\noindent N.A. Inogamov, M. Tricottet, A.M. Oparin, and S. Bouquet, 2001a,
Three-Dimensional Morphology of Vortex Interfaces
Driven by Rayleigh-Taylor or Richtmyer-Meshkov Instability,
arXiv:physics/0104084 (2001).

\noindent N.A. Inogamov, A.M. Oparin, A.Yu. Dem'yanov, L.N. Dembitskii, and V.A. Khokhlov, 2001b,
On stochastic mixing caused by the Rayleigh-Taylor instability,
Journal of Experimental and Theoretical Physics, vol. 92(4), 715-743 (2001).

\noindent N.A. Inogamov, 2002,
Statistics of long wavelength fluctuations and expansion law
for turbulent mixing zone driven by Richtmyer-Meshkov instability,
Pis'mav ZhETF (JETP Letters), vol. 75(11), 664-668 (2002).

\noindent Yu.A. Kucherenko, S.I. Balabin, R.I. Ardashova, O.E. Kozelkov, A.V. Dulov, and I.A. Romanov,
2004,
On possibility of experimental determination of integral characteristics of molecular component of mixing
by means of X-ray technique,
9th International Workshop on the Physics of Compressible Turbulent Mixing,
Cambridge, UK, 19-23 July 2004, Abstracts, p. 57;
http://www.damtp.cam.ac.uk/iwpctm9/

\noindent Yu.A. Kucherenko, A.P. Pylaev, V.D. Murzakov, A.V. Belomestnih, V.N. Popov, and A.A. Tyaktev,
2004,
On possibility of experimental determination of molecular component of non-stationary turbulent mixing,
9th International Workshop on the Physics of Compressible Turbulent Mixing,
Cambridge, UK, 19-23 July 2004, Abstracts, p. 58;
http://www.damtp.cam.ac.uk/iwpctm9/

\noindent P.F. Linden, J.M. Redondo, and D. Youngs, 1994,
Molecular mixing in Rayleigh-Taylor instability,
Journal of Fluid Mechanics, vol. 265, 97-124 (1994).

\noindent J.M. Redondo and G. Garzon, 2004,
Multifractal structure and intermittent mixing in Rayleigh-Taylor driven fronts,
9th International Workshop on the Physics of Compressible Turbulent Mixing,
Cambridge, UK, 19-23 July 2004, Abstracts, p. 91;
http://www.damtp.cam.ac.uk/iwpctm9/

\noindent O. Schilling, M. Latini, W.-S. Don, and B. Bihari, 2004,
Investigation of the large-scale and statistical properties
of Richtmyer-Meshkov instability-induced mixing,
9th International Workshop on the Physics of Compressible Turbulent Mixing,
Cambridge, UK, 19-23 July 2004, Abstracts, p. 97;
http://www.damtp.cam.ac.uk/iwpctm9/

\noindent D. Shvarts, D. Oron, D. Kartoon, et. al., 2000,
Scaling laws of nonlinear Rayleigh-Taylor and Richtmyer-Meshkov instabilities
in two and three dimensions,
C.R. Acad. Sci. Paris, t. 1, Serie IV, 719-726 (2000).

\noindent J.G. Wouchuk, 2001a,
Growth rate of the linear Richtmyer-Meshkov instability
when a shock is reflected,
Phys. Rev. E, vol. 63, 056303 (2001a).

\noindent J.G. Wouchuk, 2001b,
Growth rate of the Richtmyer-Meshkov instability
when a rarefaction is reflected,
Phys. of Plasmas, vol. 8(6), 2890-2907 (2001b).

\noindent D.L. Youngs, 2004,
Effect of initial conditions on self-similar turbulent mixing,
9th International Workshop on the Physics of Compressible Turbulent Mixing,
Cambridge, UK, 19-23 July 2004, Abstracts, p. 123;
http://www.damtp.cam.ac.uk/iwpctm9/

\newpage
\centerline{{\bf Figures}}

\begin{center}
\includegraphics[width=12cm]{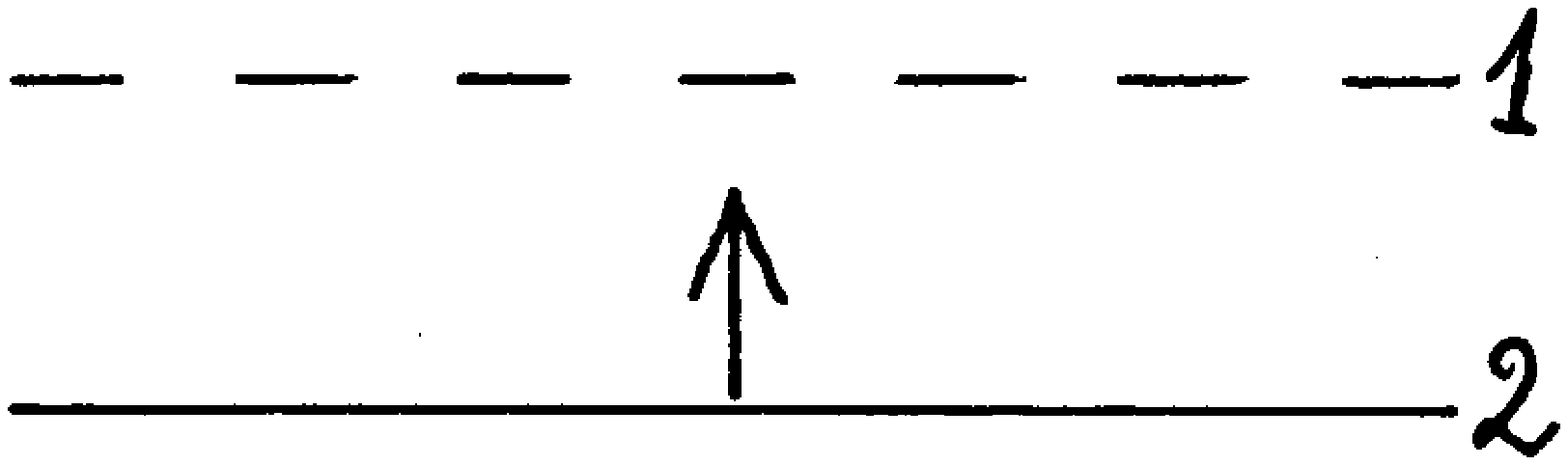}

Fig. 1. Unperturbed flow.
It is shown
approach of a shock wave (a straight line 2 and an arrow)
to contact boundary (a straight line 1).
\end{center}

\begin{center}
\includegraphics[width=12cm]{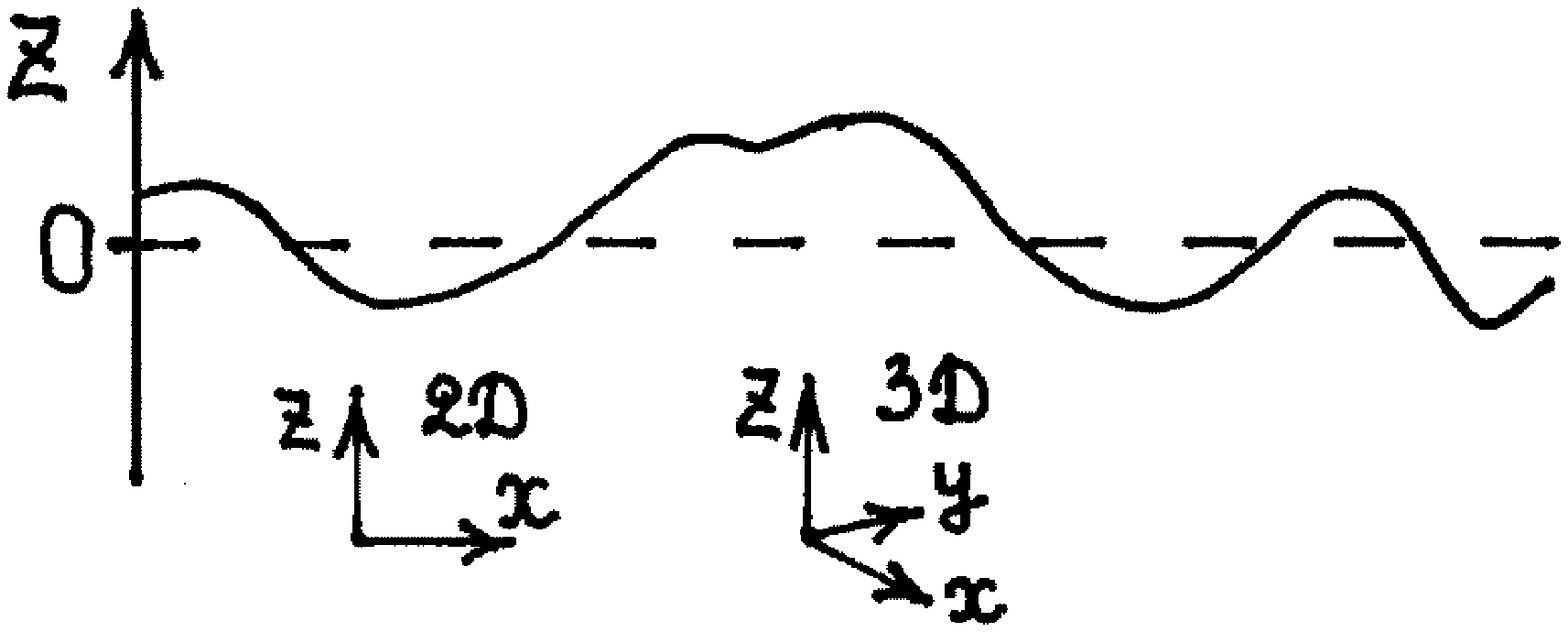}

Fig. 2. Richtmyer-Meshkov instability
arises from surface perturbations
(a continuous curve).
Unpertubed position of boundary is shown by dashed straight line.
Definition of longitudinal and transverse coordinates.
\end{center}

\begin{center}
\includegraphics[width=12cm]{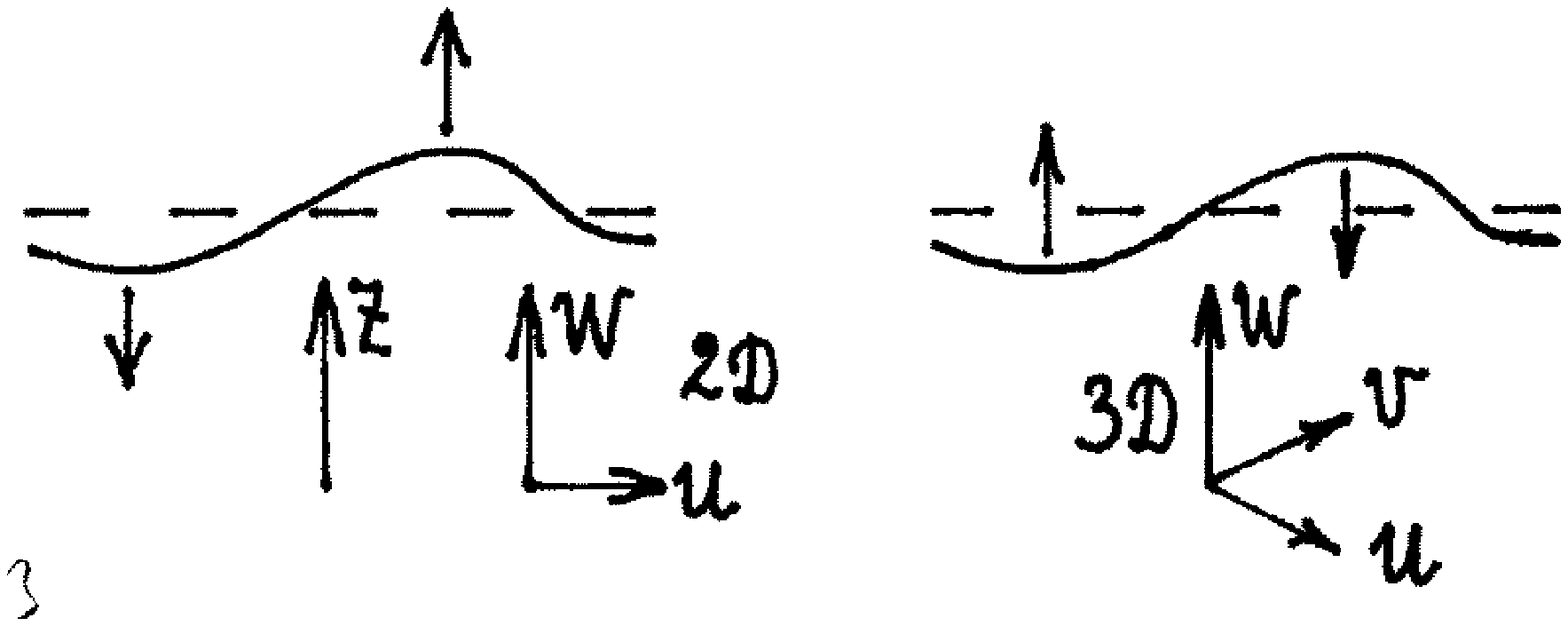}

\end{center}

Fig. 3. Baroclinic "processing" by a shock wave
of surface perturbations $ \eta $
into velocity field $ \vec v. $
The form (continuous curves)
and directions of movement (arrows) of contact boundary
after passage of a shock for cases in a phase (at the left)
and in a counterphase (on the right).
Shaped straight line is position of unperturbed boundary.
\newpage

\begin{center}
\includegraphics[width=12cm]{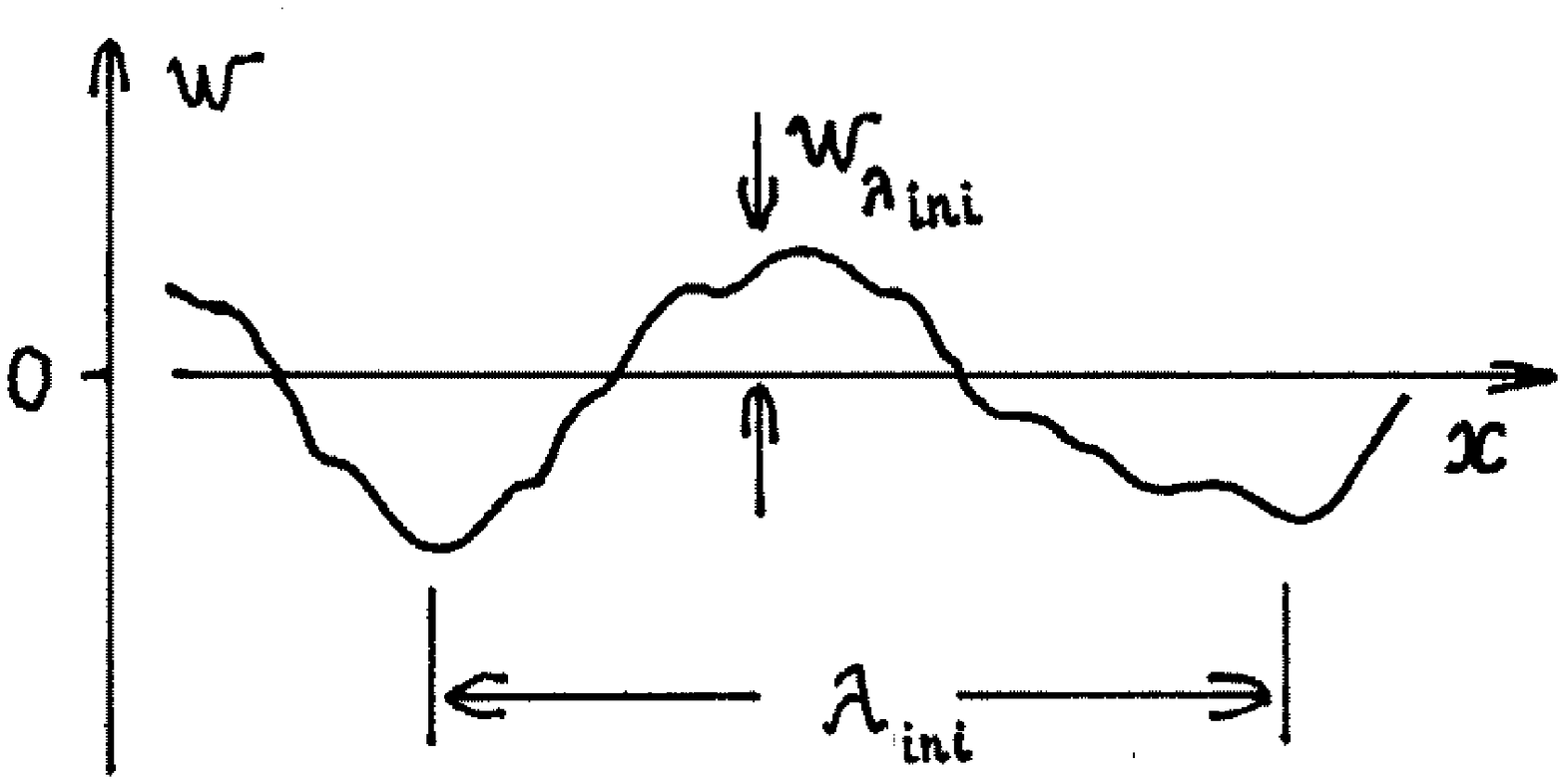}

Fig. 4. The most interesting example
(by virtue of the typicalness)
of the general stochastic function
set in scales $ \lambda_{ini} $ and $w_{\lambda ini} $
$(w $ is a component of velocity on an axis $z). $
The sign-variable field of velocity
in the near boundary layer
arising after passage of a shock
through the perturbed boundary is presented.
\end{center}

\begin{center}
\includegraphics[width=12cm]{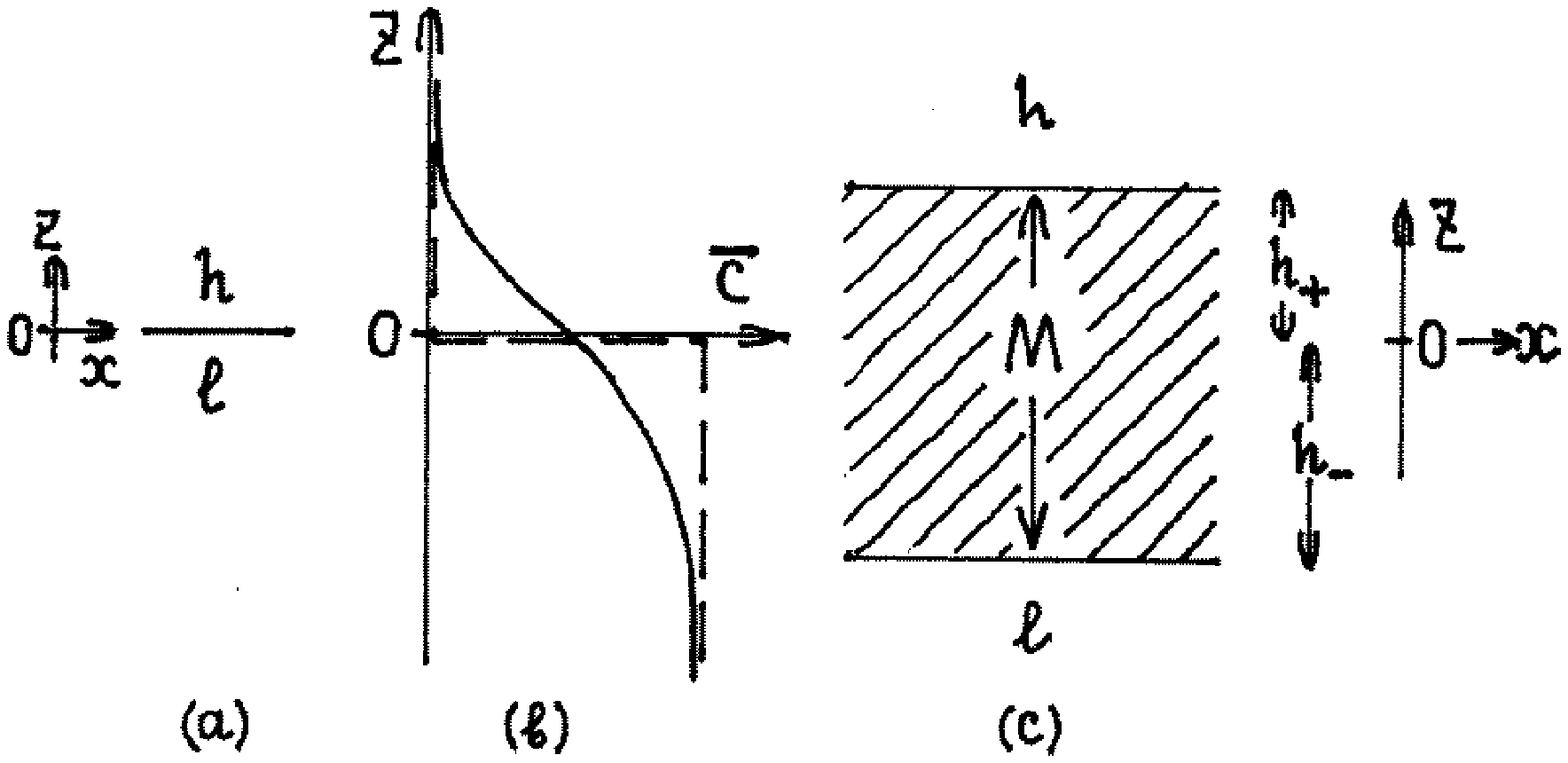}

Fig. 5. The average distribution of light "l" and heavy "h" phases
(a) prior to the beginning of mixing
and (c) after the beginning of mixing.
As a result of interpenetration of phases
intermediate layer "M" with a mixture of "l" and "h" phases is formed.
\end{center}

\begin{center}
\includegraphics[width=11cm]{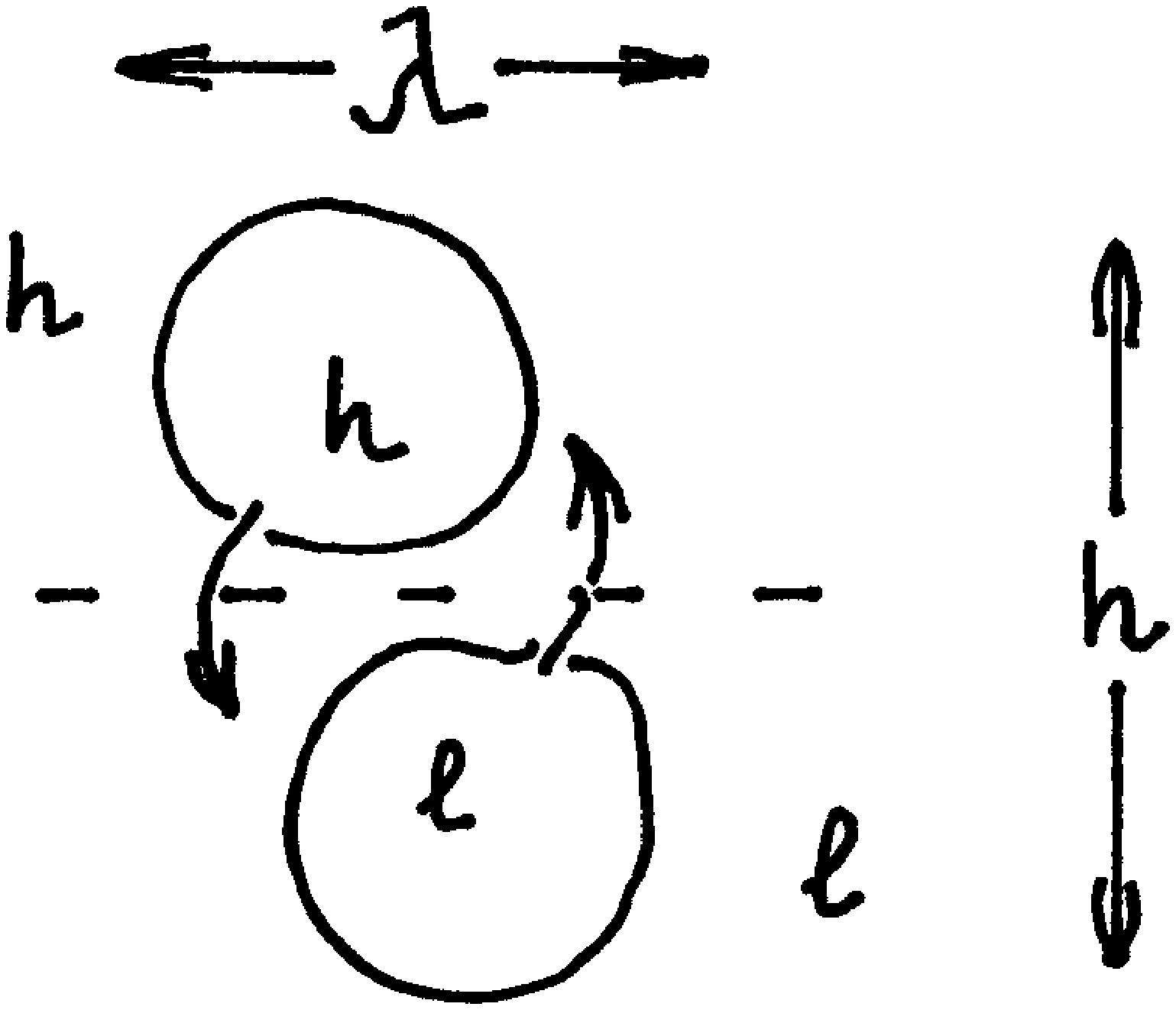}

Fig. 6. Involving into turbulent movement of volumes of light "l"
and heavy "h" liquids.
\end{center}

\begin{center}
\includegraphics[width=12cm]{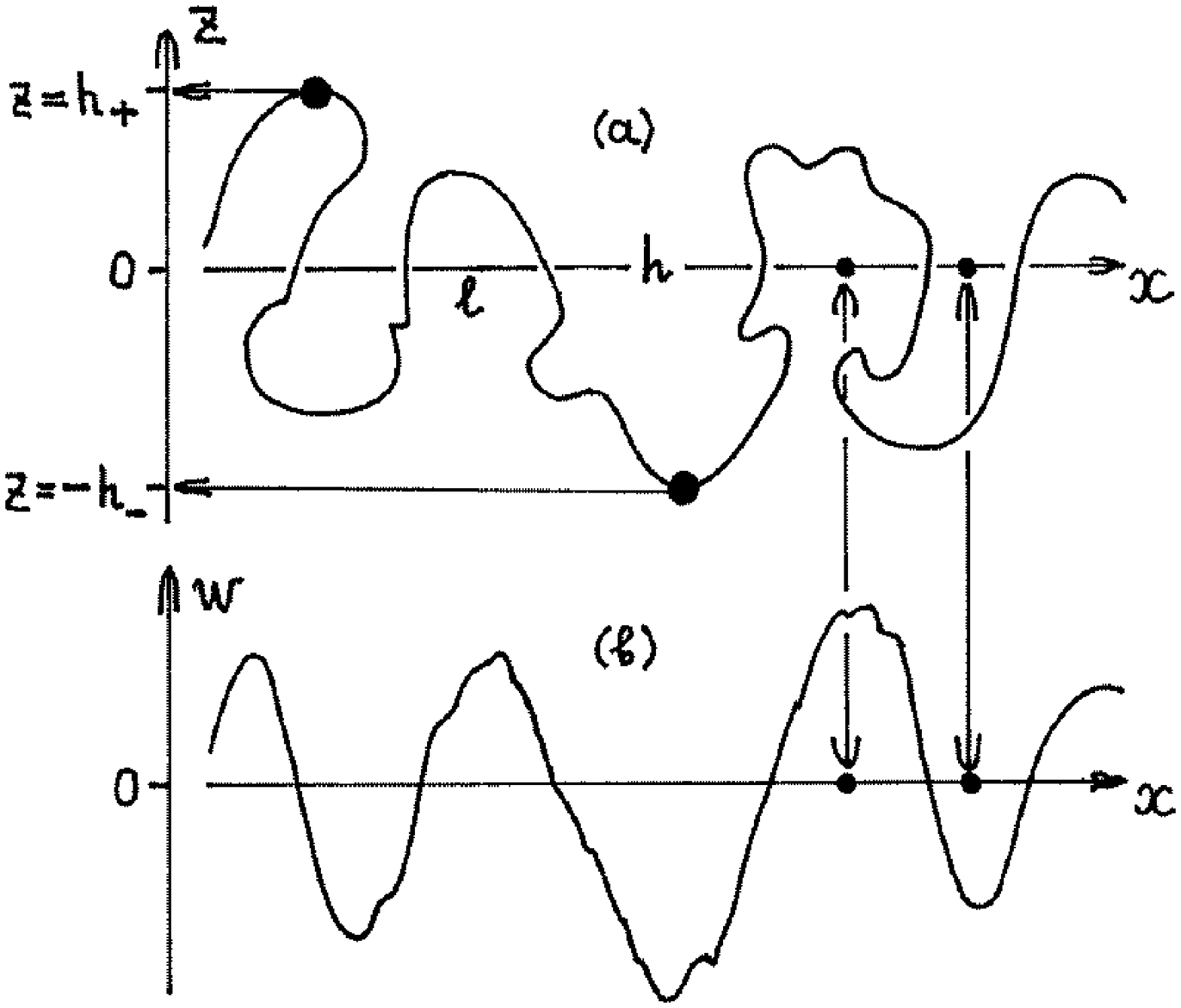}

Fig. 7. (a) The instant picture alternating "l" and "h" columns.
(b) Sign variability of velocity $w(x, y_{fix}, z=0, t_{fix}). $
The plane $z=0 $ corresponds to position of unperturbed contact.
\end{center}

\begin{center}
\includegraphics[width=12cm]{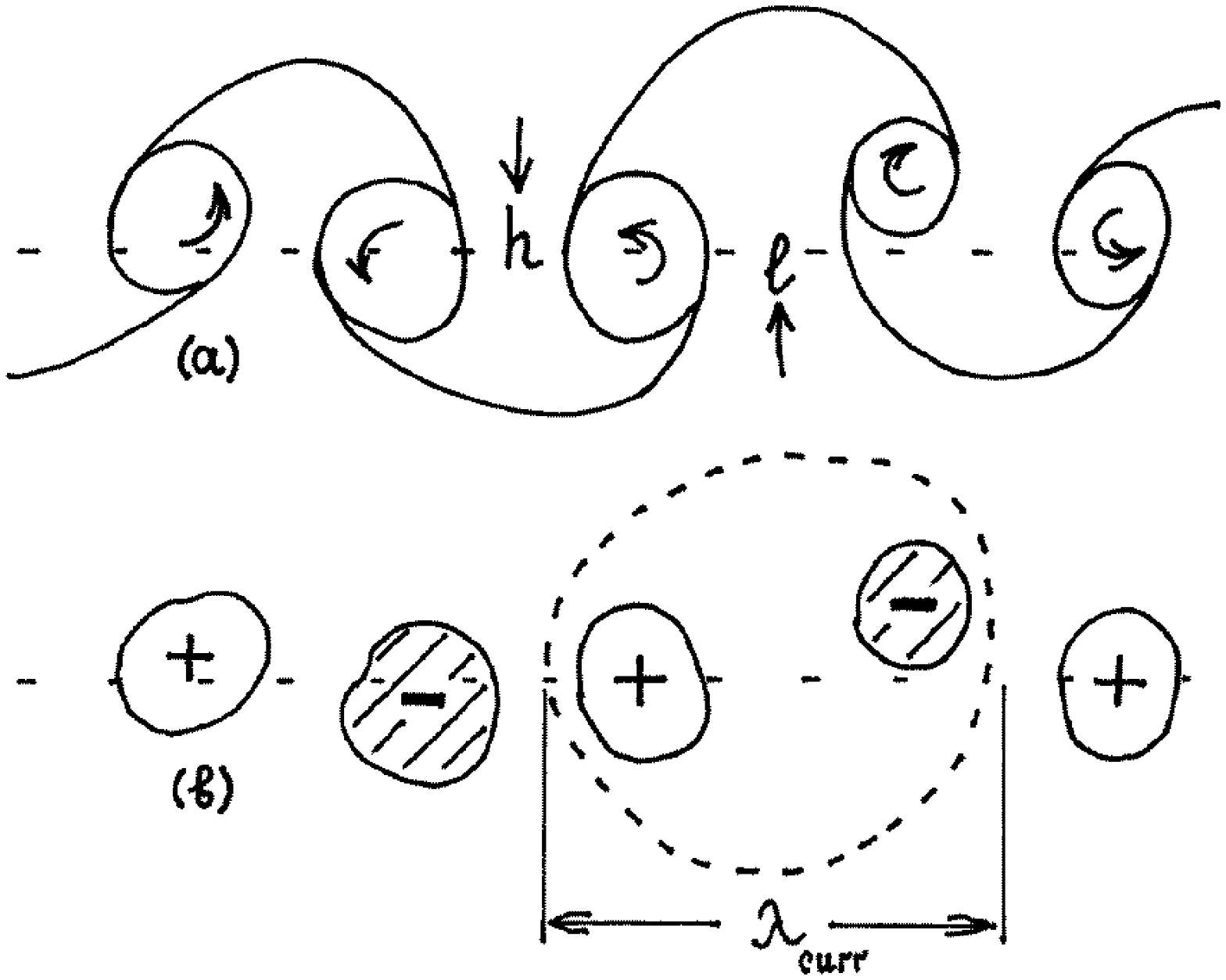}

Fig. 8. Large-scale structure and a vortical chain.
(a) "Blowing" of "l" and "h" columns by pairs of vortical spots.
(b) A chain of large-scale sign-variable vorticities
or a chain of dipoles $ \lambda\times h $ from pairs of vorticities.
\end{center}

\begin{center}
\includegraphics[width=12cm]{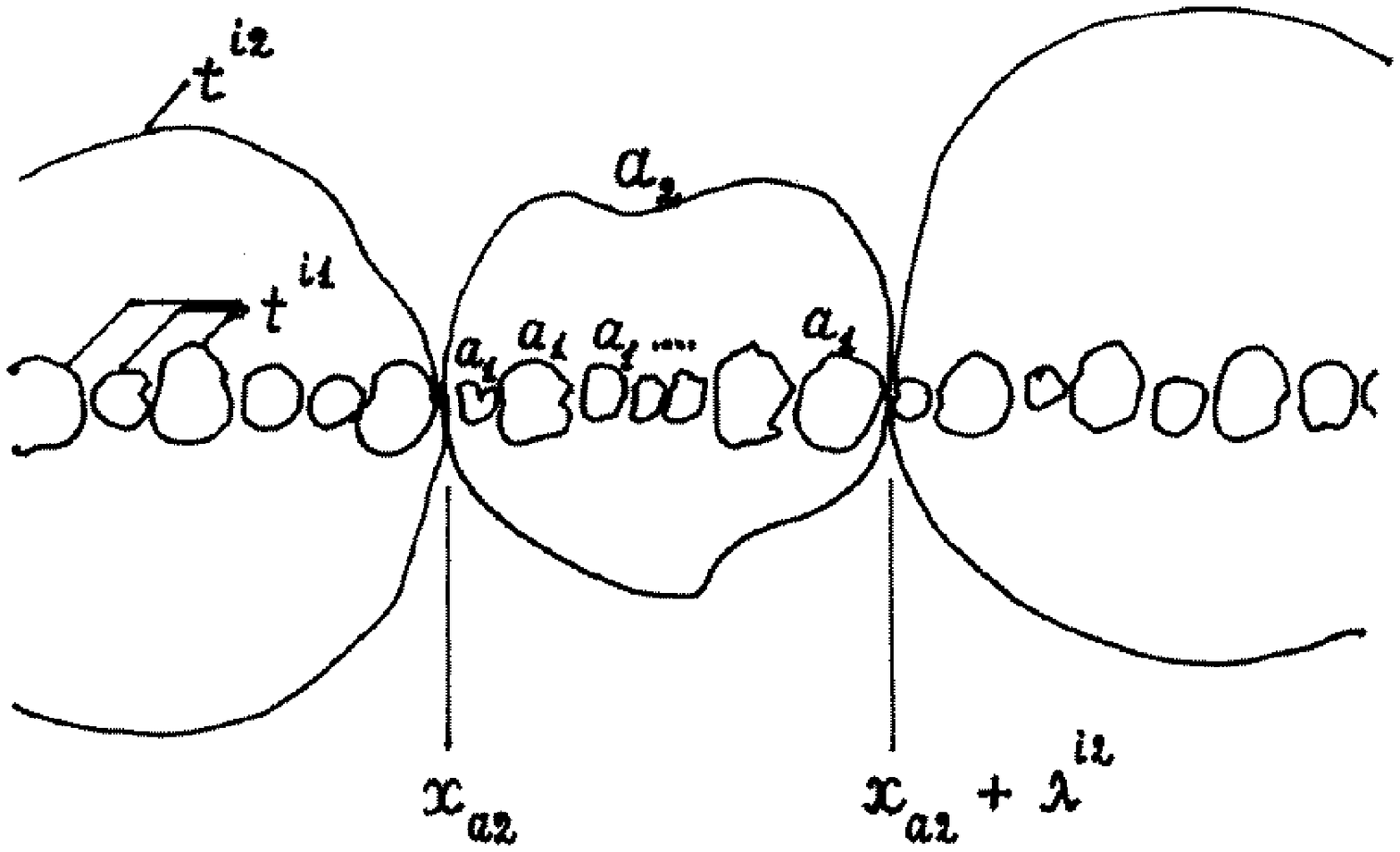}

Fig. 9. Formation of a claster dipole
from small dipoles of the previous stages.
\end{center}

\begin{center}
\includegraphics[width=12cm]{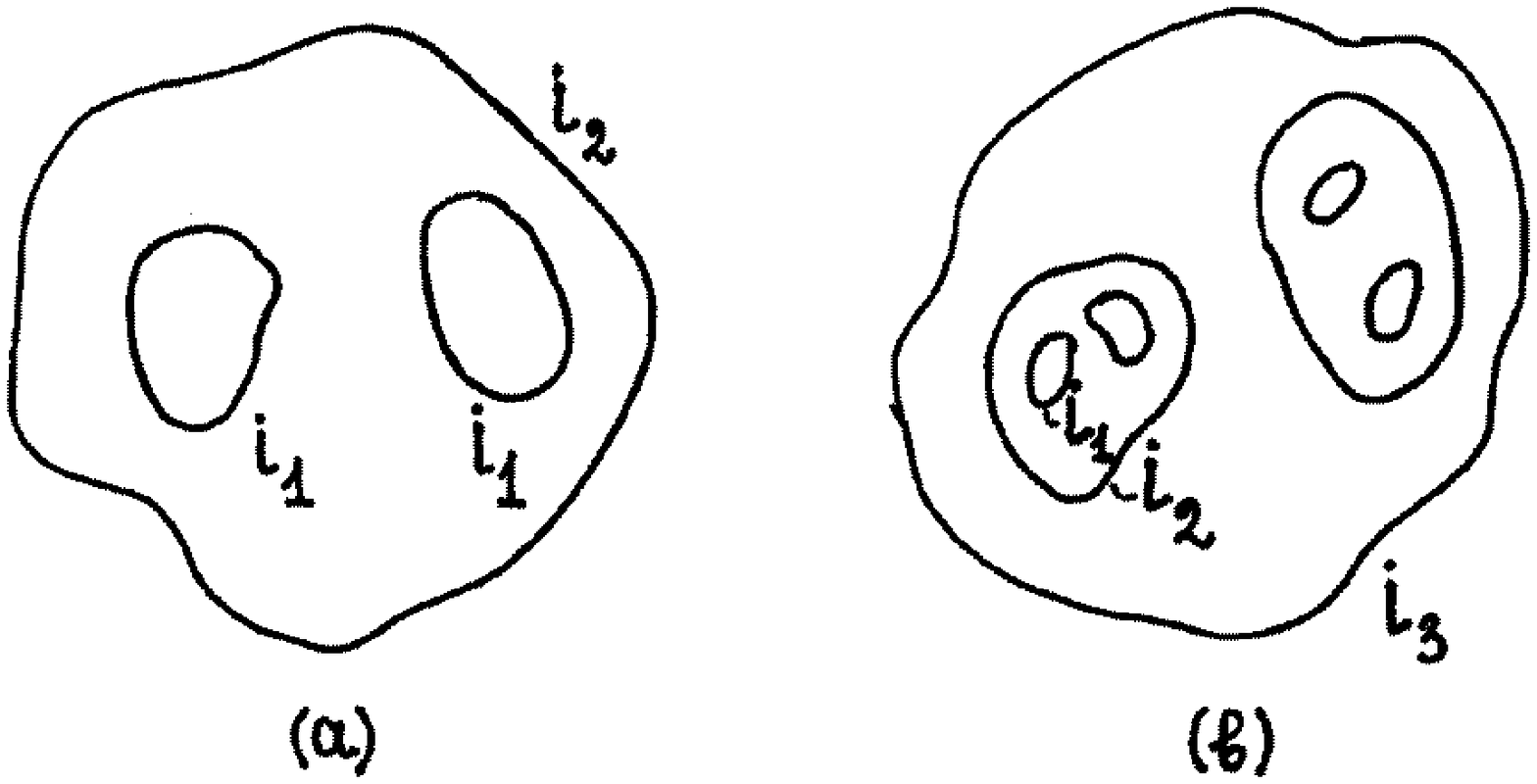}

Fig. 10. Compound character of large-scale dipoles at late stages of mixing.
\end{center}

\begin{center}
\includegraphics[width=12cm]{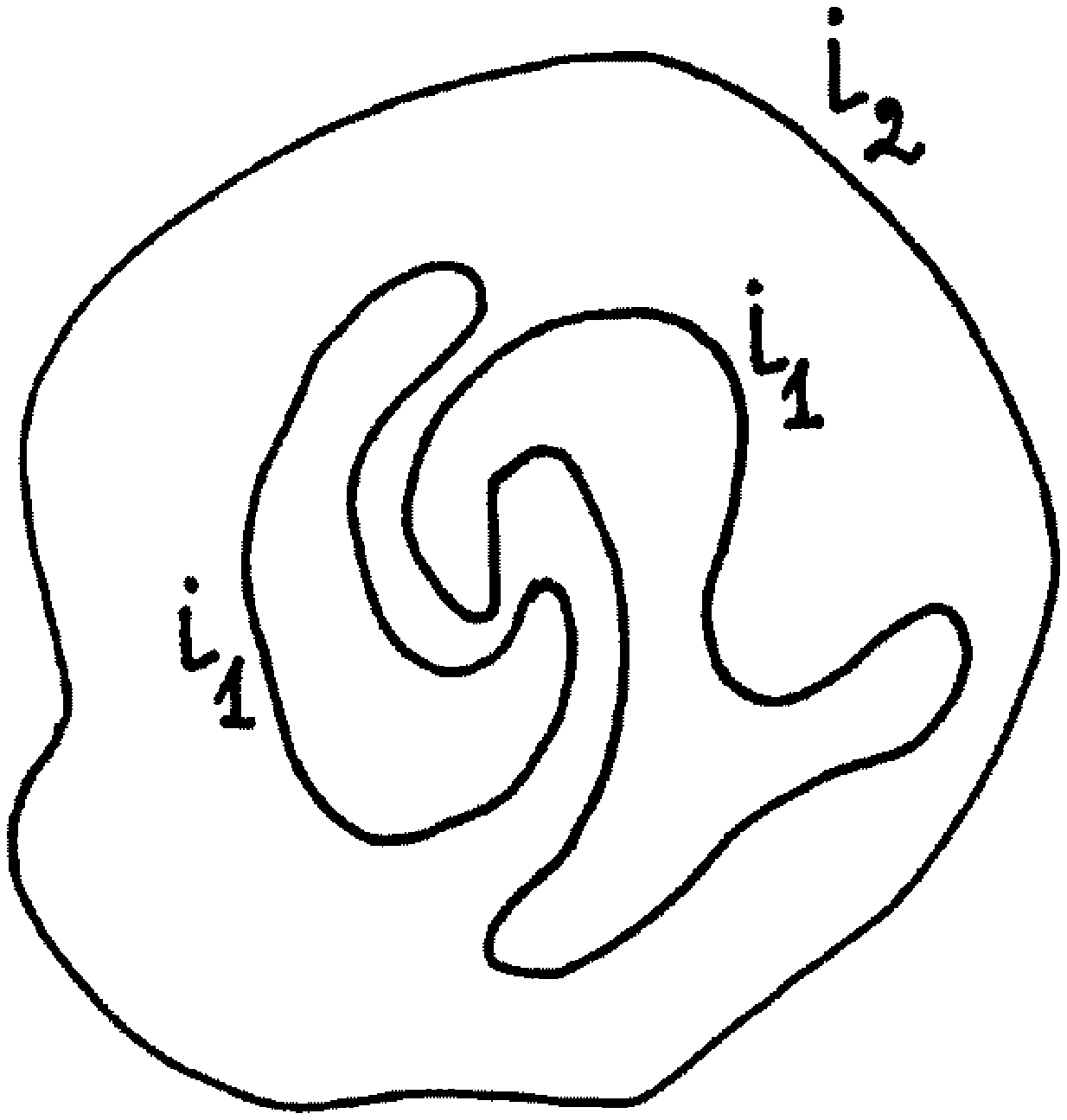}

Fig. 11. Extension and deformation of dipoles of generation $i_1 $
inside dipoles of the subsequent generation $i_2. $
\end{center}

\begin{center}
\includegraphics[width=12cm]{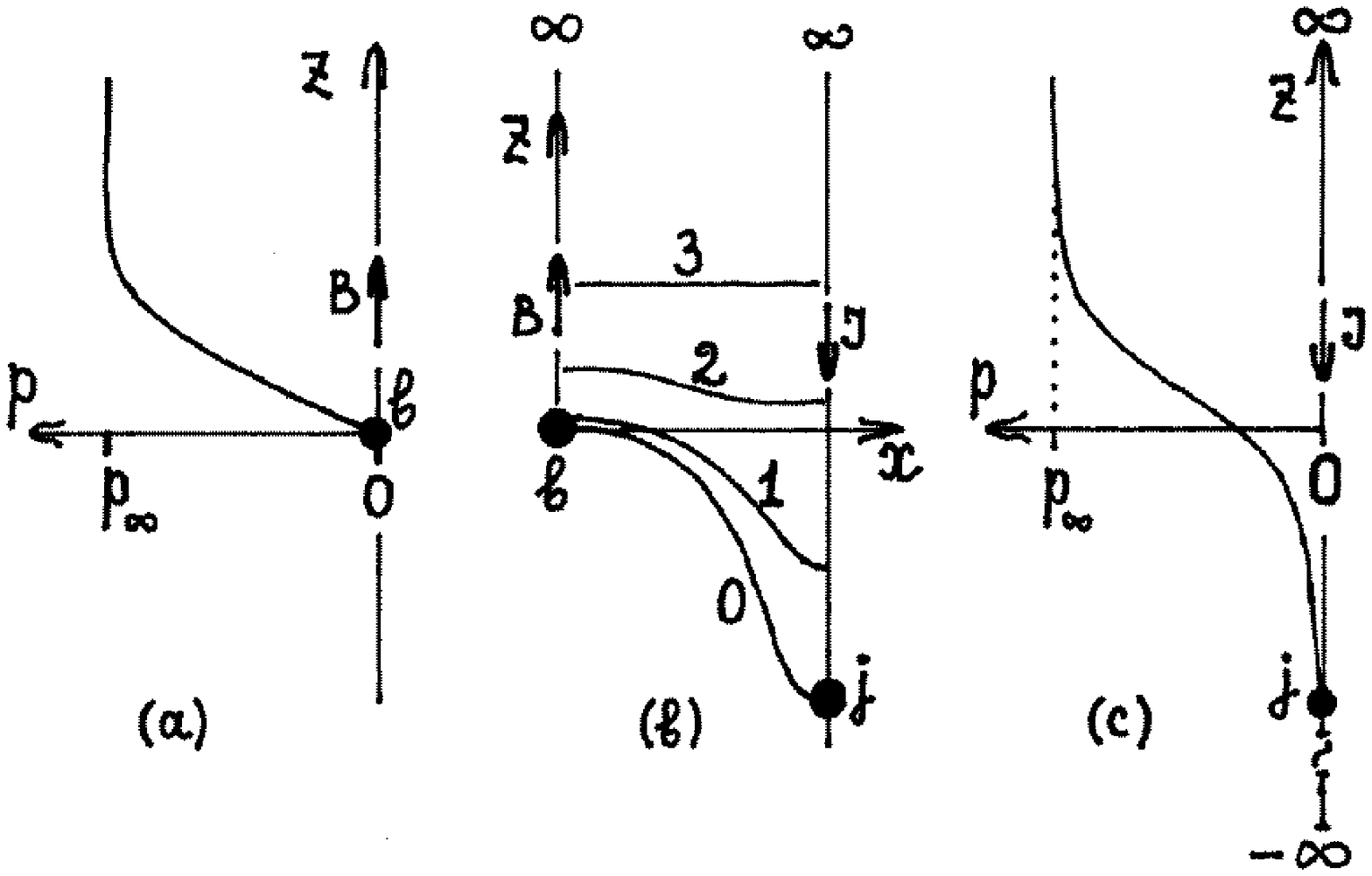}

Fig. 12. Squeezing $ (p_{\infty}> 0) $ of bulk $ (+ \infty) $ of liquids
by ram pressure of bubbles pressed into it.
(a) $p(z) $ on a straight line (b$\infty); $
(b) Increase of $p $ on isobars 0 (0 is a boundary of a bubble "bj"), 1, 2 and 3.
At boundary "bj" $p_{CB} =0. $
(c) $p(z) $ on (j$\infty). $

\end{center}

\begin{center}
\includegraphics[width=12cm]{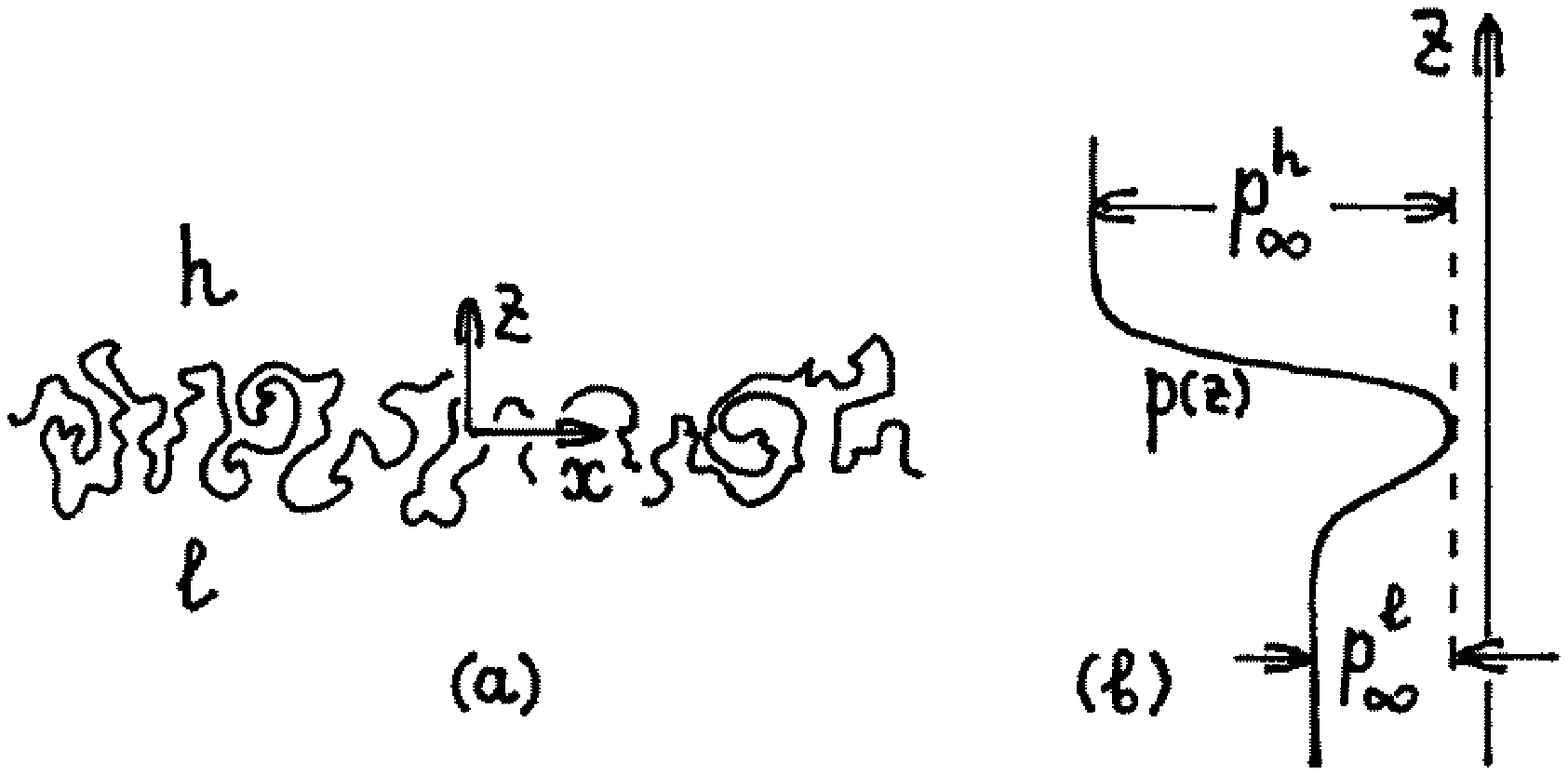}

Fig. 13. (a) The layer with the turbulent motion
concentrated near a plane $z=0. $
(b) A profile $p(z) $ with "hole" $ (\mu\neq 0) $ in the region of movement.
\end{center}

\begin{center}
\includegraphics[width=11.5cm]{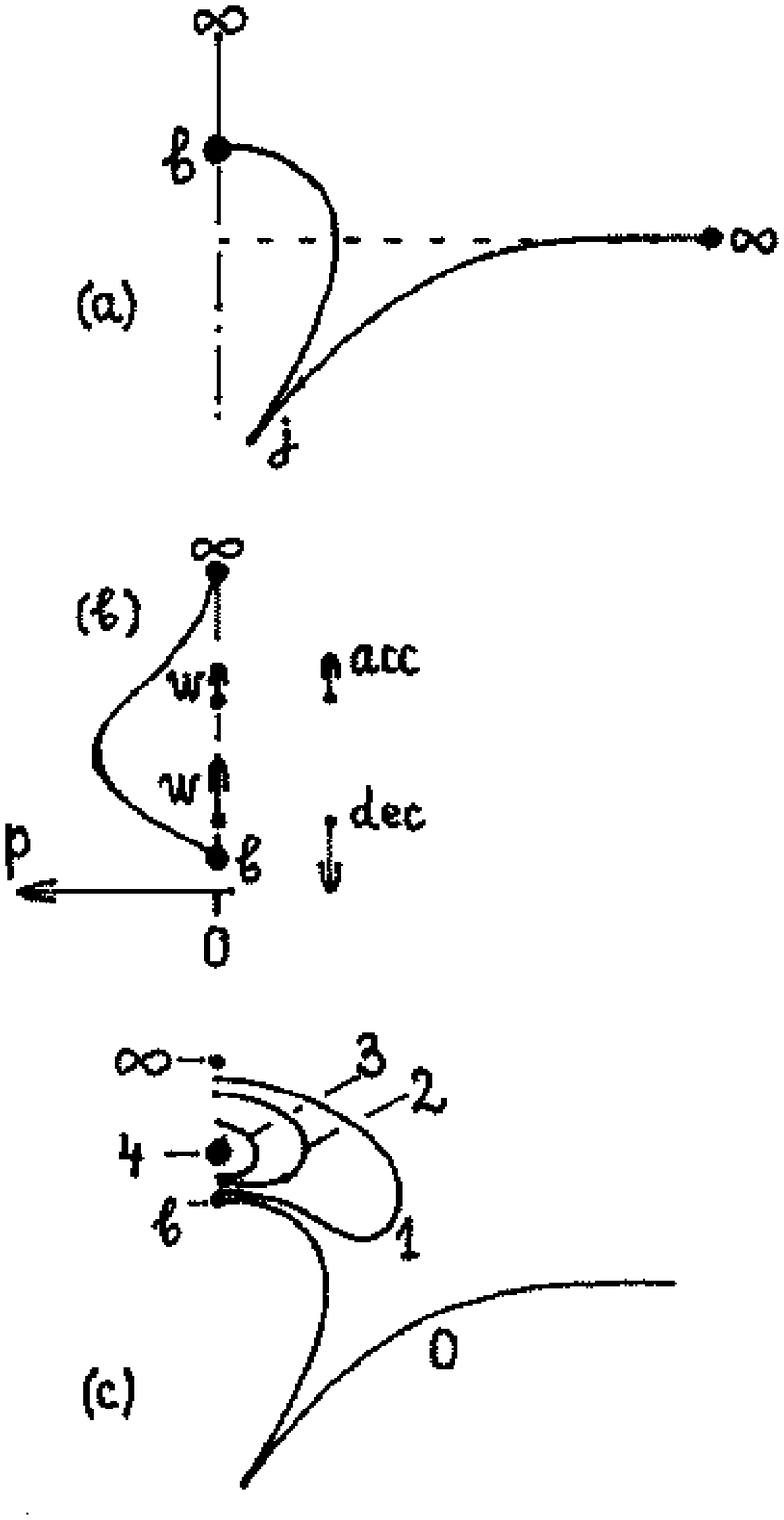}

Fig. 14. (a) Penetration of bubble. Geometry of flow.
(b) Distribution $p $ lengthways (b$\infty). $
(c) An instant field of pressure.
\end{center}

\begin{center}
\includegraphics[width=12cm]{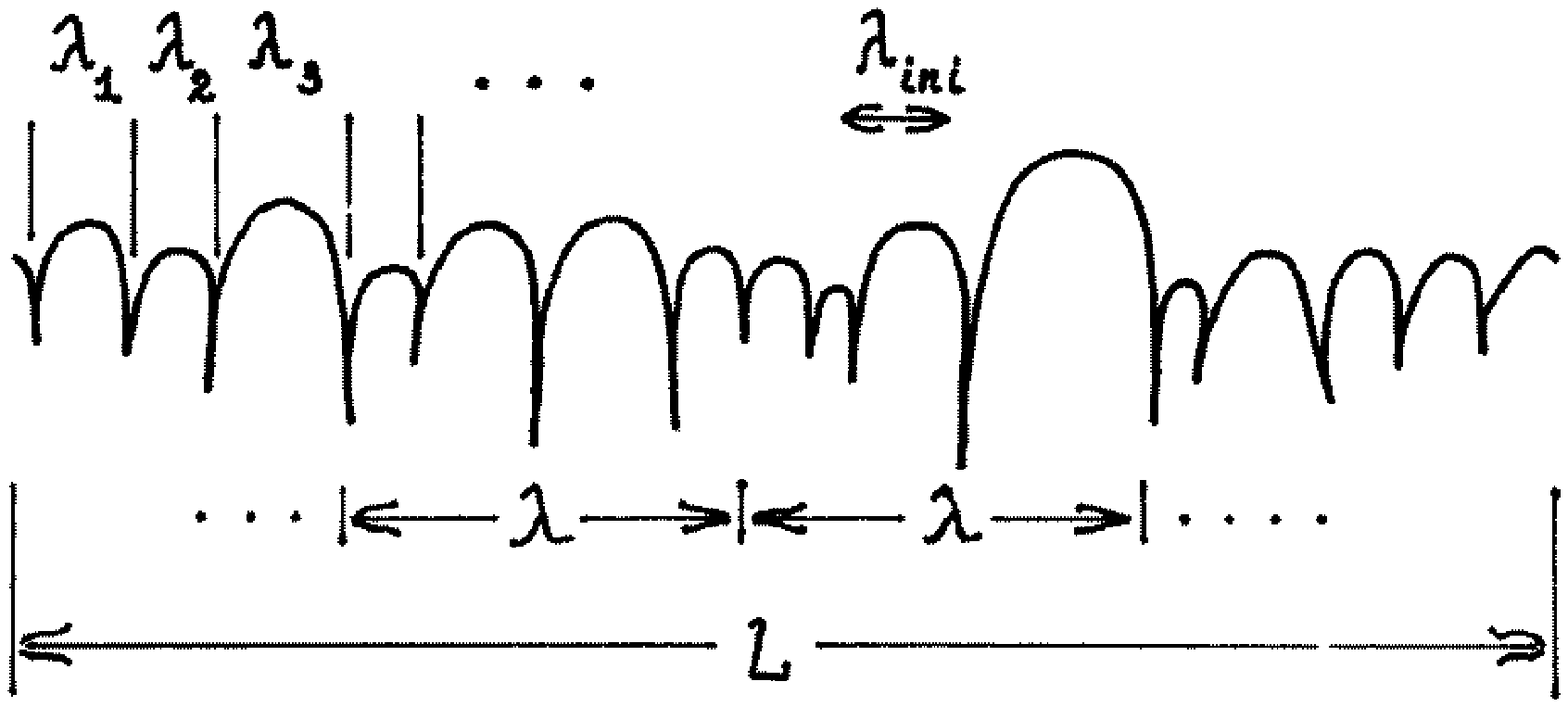}

Fig. 15. Ensemble of bubbles $\lambda_i $
with the average size $ \lambda_{ini} $
and groups of bubbles, got on pieces $ \lambda $
($\lambda$-groups).
\end{center}

\begin{center}
\includegraphics[width=12cm]{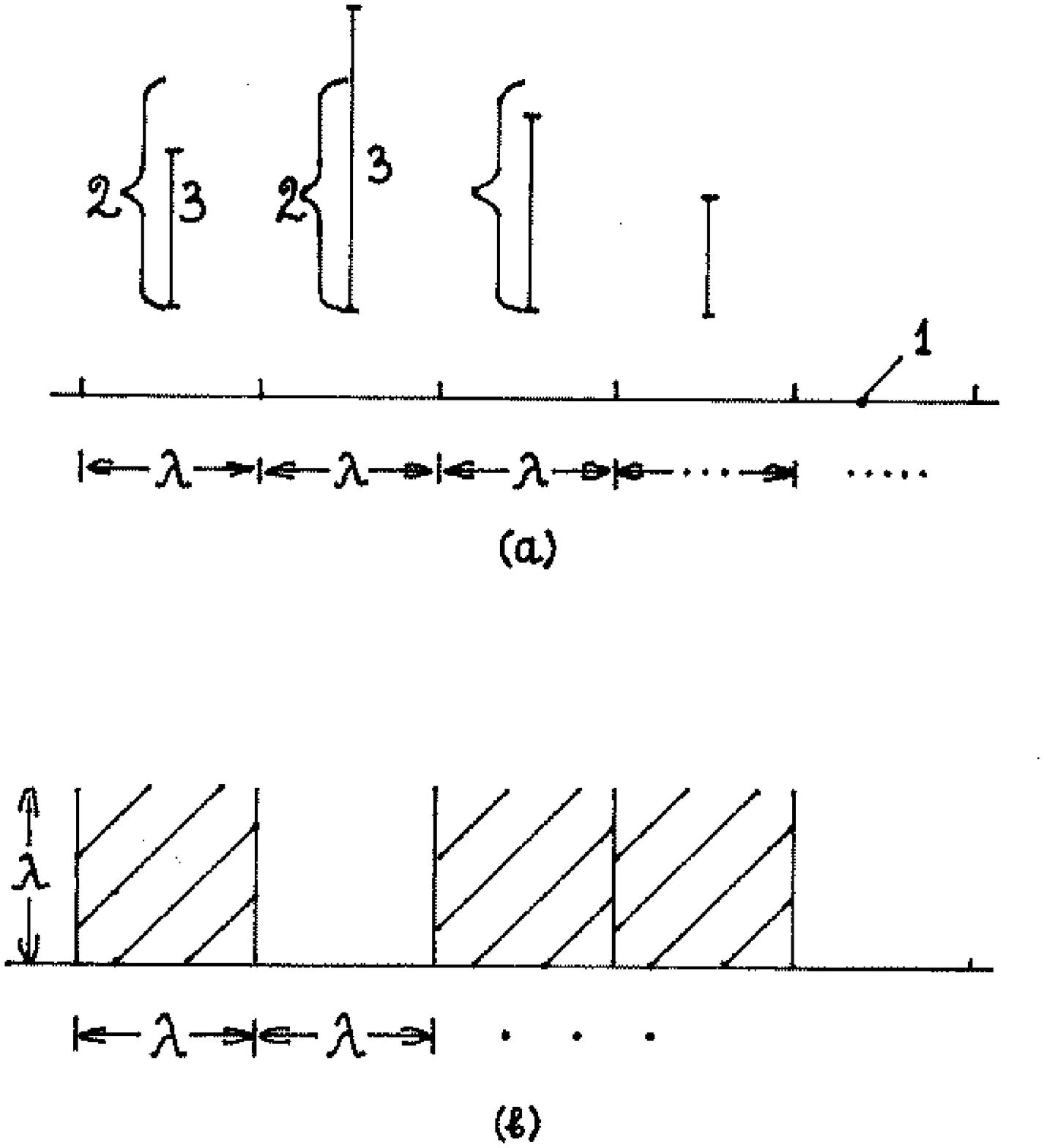}

Fig. 16. (a) Changeability of pressure (pieces 3),
created by $"\lambda$-groups" of bubbles.
(b) Random alternation of squares $ \lambda\times\lambda $
with positive and negative (are shaded) increments of velocity
$w_{\lambda}. $
\end{center}

\begin{center}
\includegraphics[width=12cm]{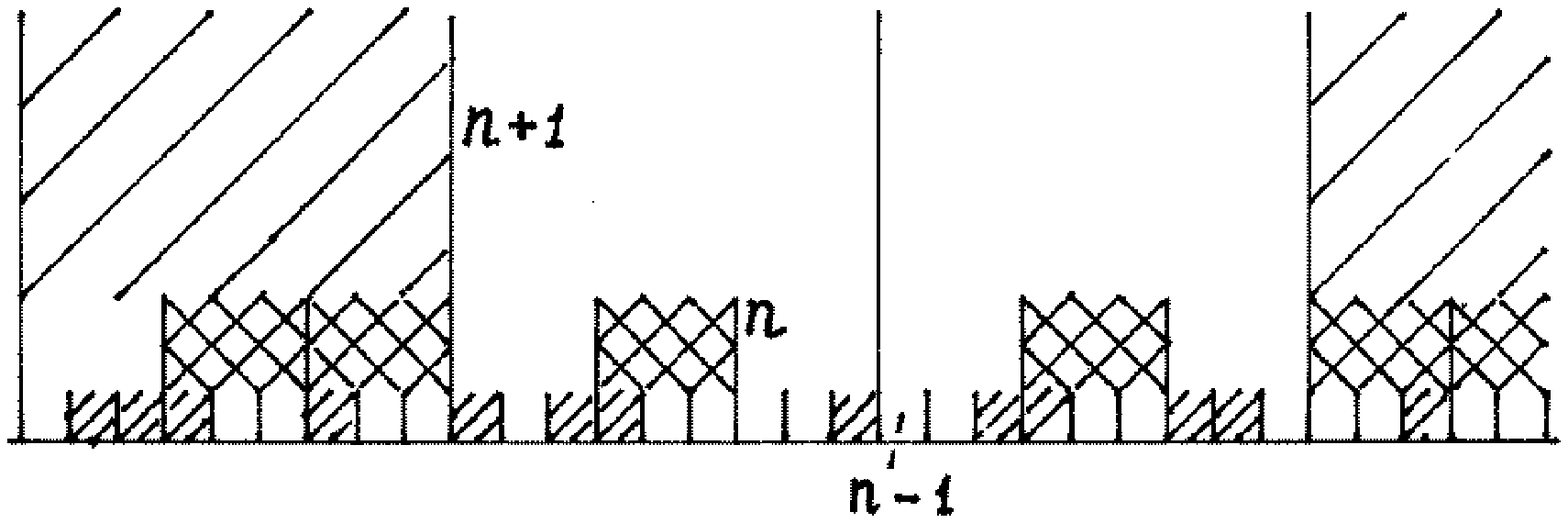}

Fig. 17. Hierarchy of changes of a sign
of velocity $w (x, z): $
small-scale changeability near surface,
large-scale - it is far in a bulk of liquid.
\end{center}

\begin{center}
\includegraphics[width=12cm]{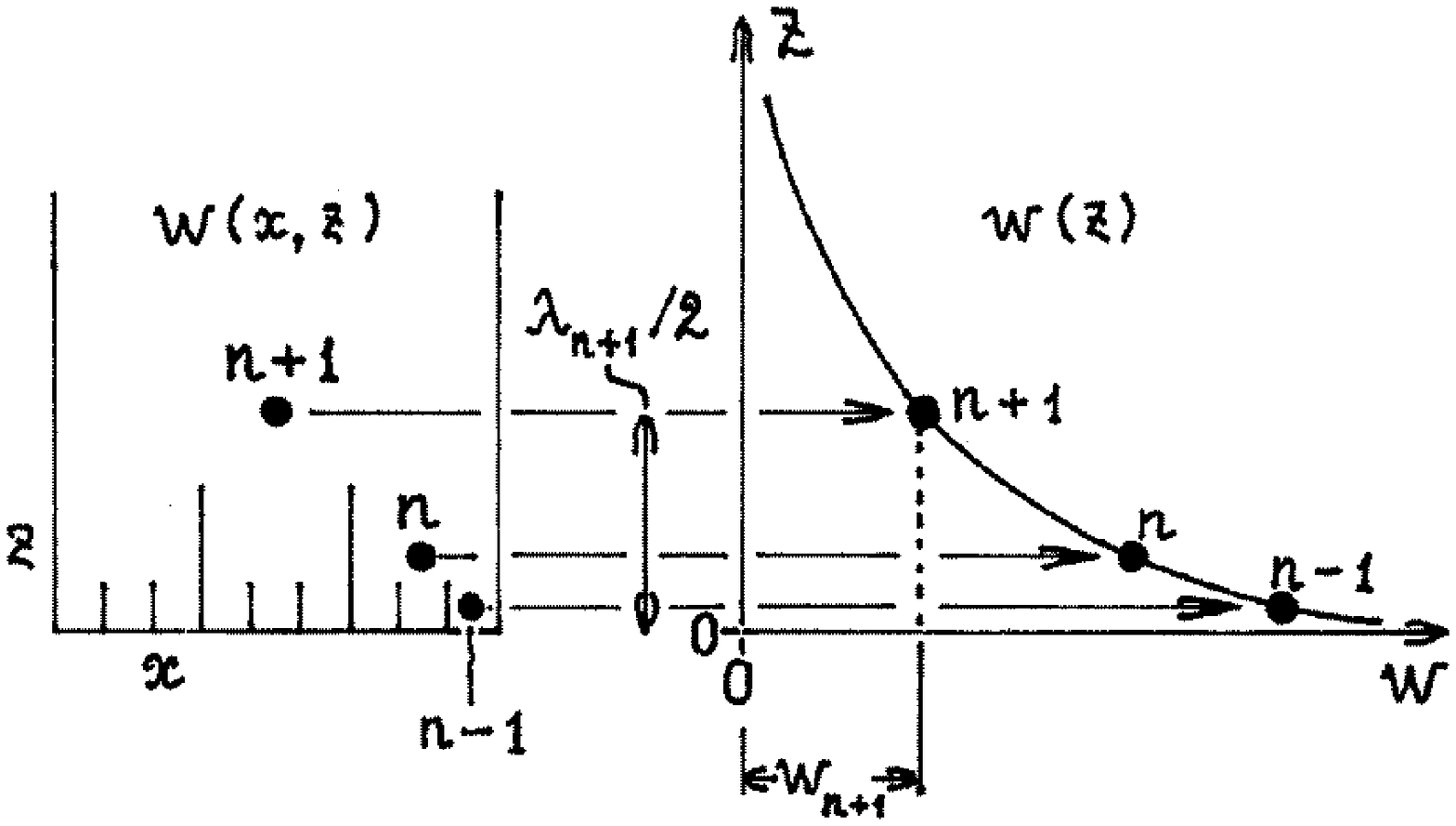}

Fig. 18. (a) Scales $n-1, n, n+1 $ and the squares connected to them.
(b) "Designing" the centers of squares from the left figure on right.
\end{center}

\begin{center}
\includegraphics[width=12cm]{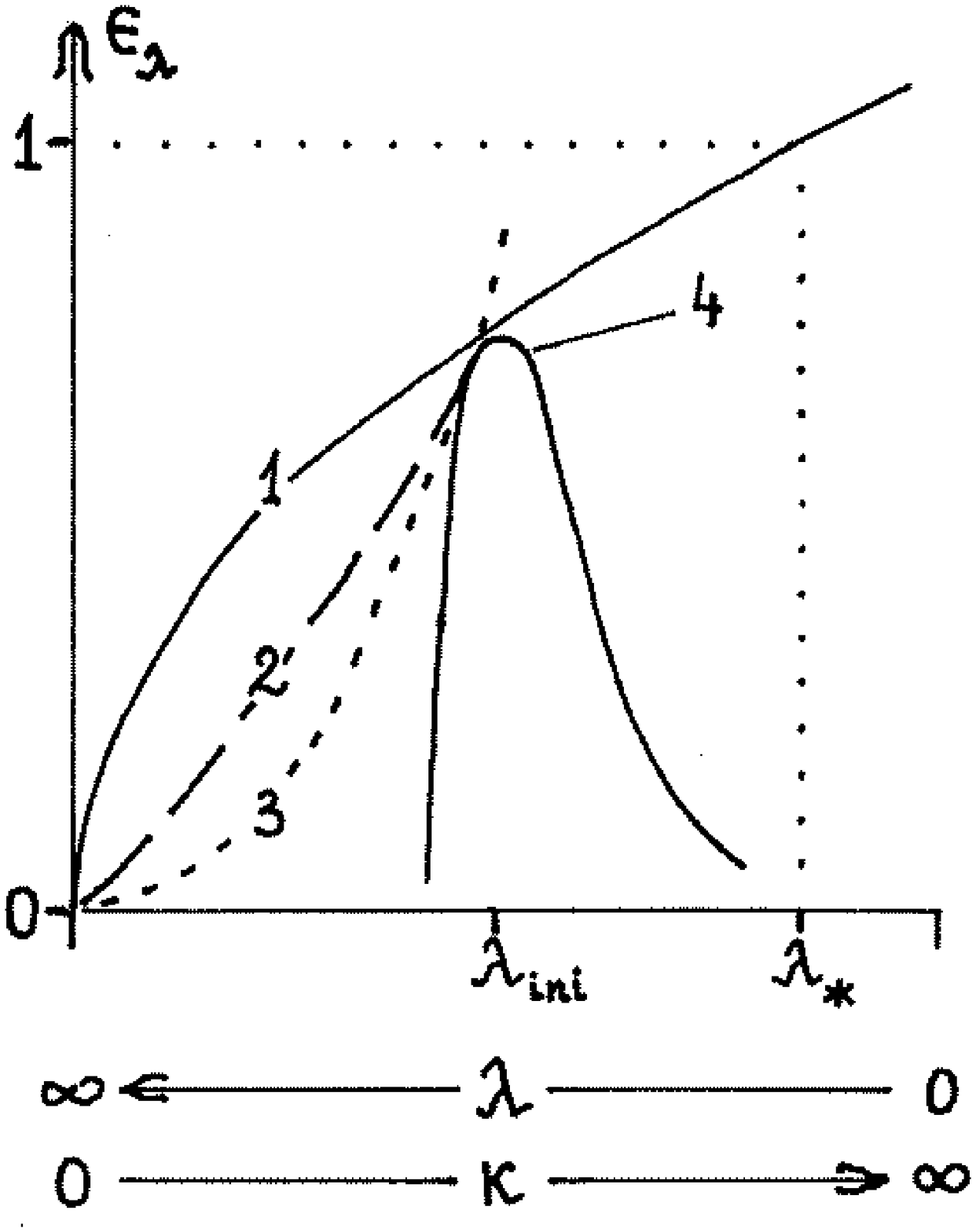}

Fig. 19. The combined perturbations
from (1) added long wavelength "noise" (a curve 1)
and (2) "located" on $ \lambda $ a spectrum (a curve 4).
Perturbations 4 generate a large-scale tail (a curve 2).
\end{center}

\newpage
\begin{center}
\includegraphics[width=12cm]{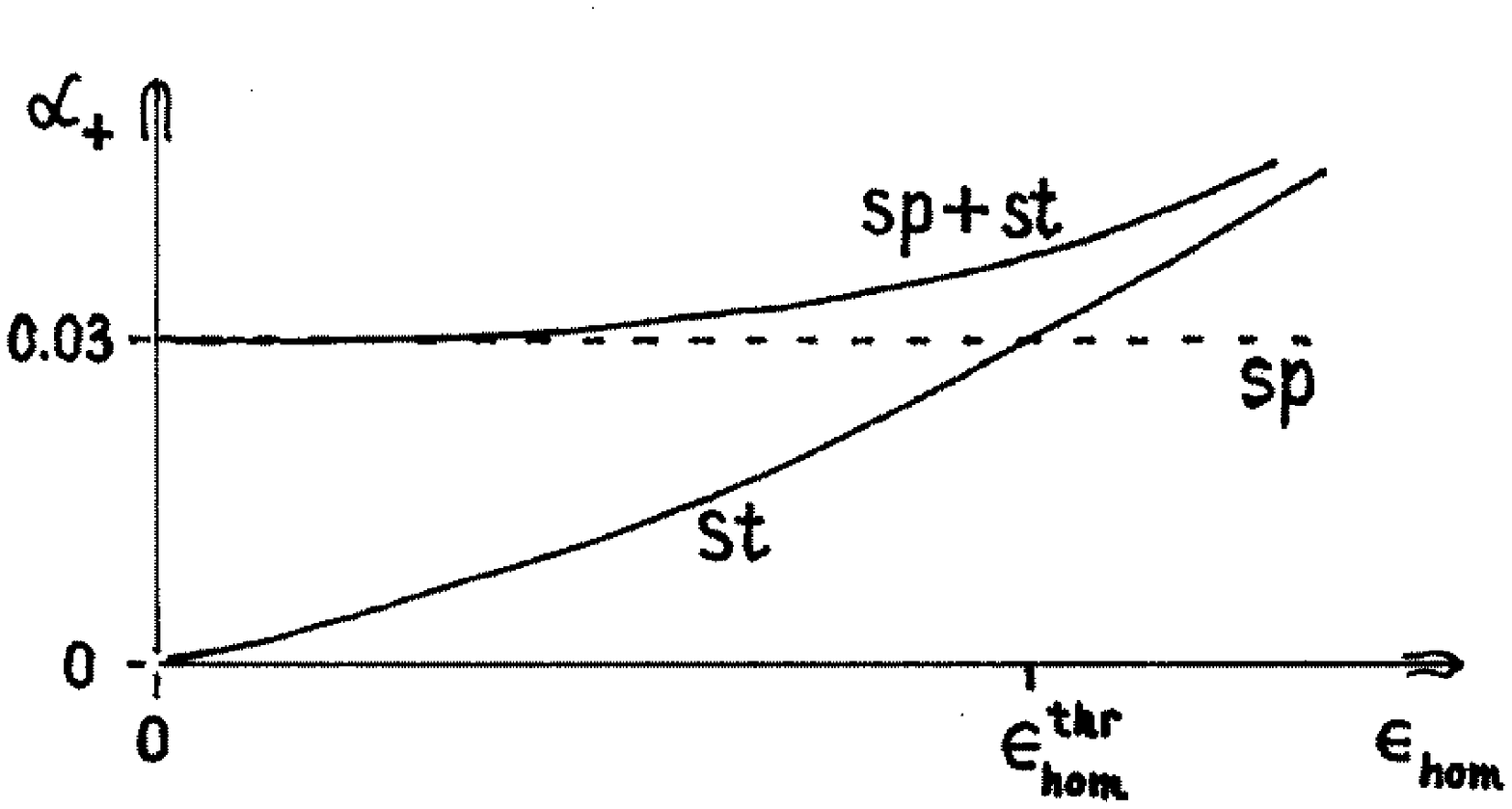}

Fig. 20. Dependence of factor $\alpha_+^{sp+st}$
for the combined perturbations
from relative amplitude of long wavelength homogeneous noise
$ \epsilon_{hom}. $
\end{center}

\begin{center}
\includegraphics[width=12cm]{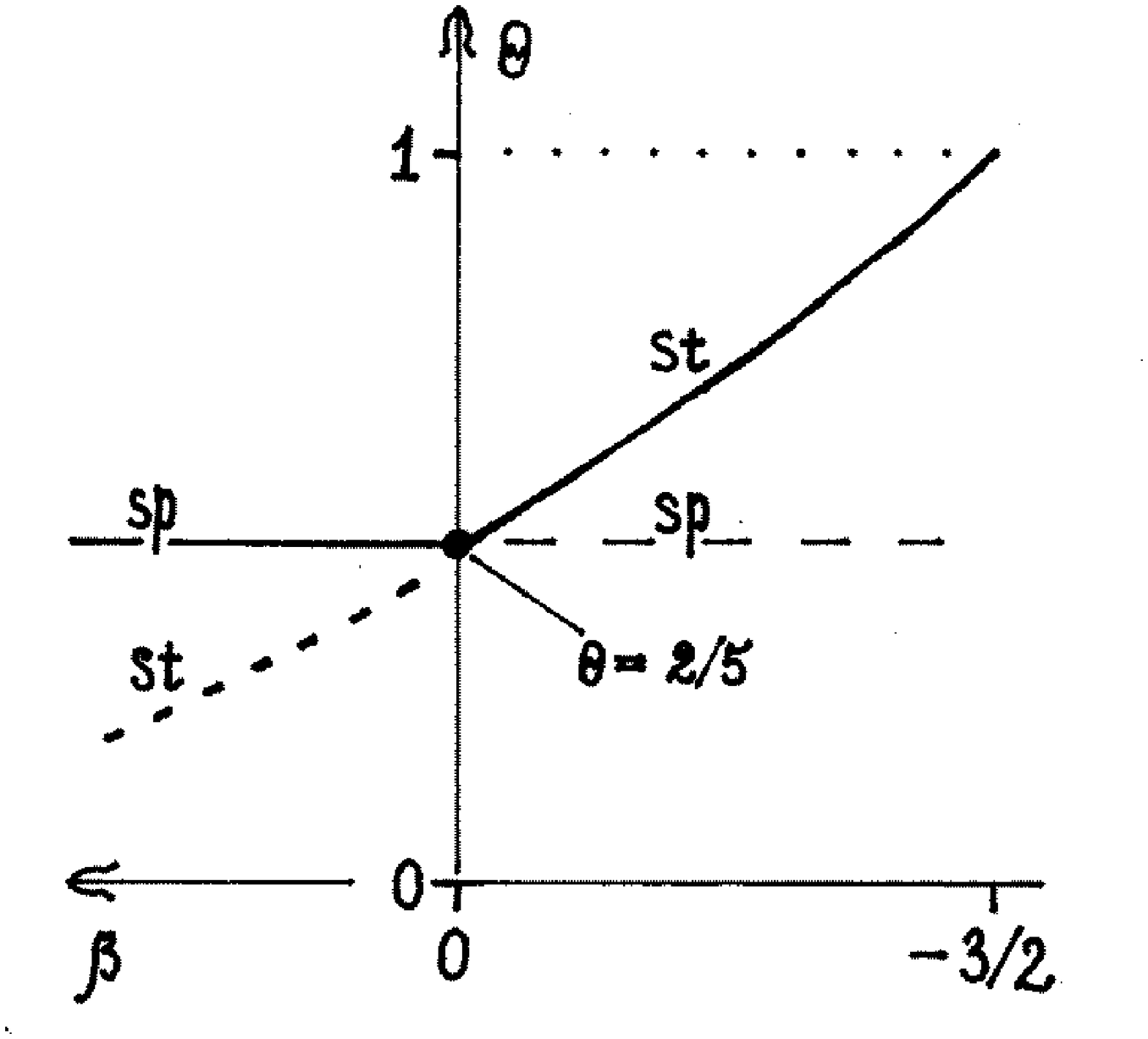}

Fig. 21. Analogue of Fig. 20 in Richtmyer-Meshkov case:
$ \theta^{sp} $ limits an interval of $ \theta $ from below.
\end{center}

\end{document}